\documentclass[Journal,letterpaper,NoLists,NoLineNumbers,InsideFigs]{ascelike-new}
\usepackage{amsmath}
\usepackage[figurename=Fig.,labelfont=bf,labelsep=period]{caption}
\usepackage{color}
\usepackage[T1]{fontenc}
\usepackage{graphicx}
\usepackage[utf8]{inputenc}
\usepackage{lmodern}
\usepackage{subcaption}
\usepackage{newtxtext,newtxmath}
\usepackage[colorlinks=true,citecolor=red,linkcolor=black]{hyperref}
\usepackage[authoryear]{natbib}
\usepackage{multirow}
\usepackage{booktabs}
\usepackage{longtable}
\usepackage{array}
\usepackage{wrapfig}
\usepackage{float}
\usepackage{colortbl}
 \usepackage{pdflscape}
 \usepackage{tabu}
 \usepackage{threeparttable}
 \usepackage{threeparttablex}
 \usepackage[normalem]{ulem}
 \usepackage{makecell}
 \usepackage{xcolor}
 \usepackage{gensymb}
 \usepackage{pdfpages}
 
\newcommand{\halfblankline}{\quad\vspace{-0.5\baselineskip}\pagebreak[3]}

\title{Full Title: A Bayesian Markov model with P{\'o}lya-Gamma sampling for estimating individual behavior transition probabilities from accelerometer classifications}

\author[1,*]{Toryn L. J. Schafer}
\author[1]{Christopher K. Wikle}
\author[2]{Jay A. VonBank}
\author[2]{Bart M. Ballard}
\author[3]{Mitch D. Weegman}

\affil[1]{Department of Statistics, University of Missouri, Middlebush Hall, Columbia, MO 65211}
\affil[2]{Caesar Kleberg Wildlife Research Institute, Texas A\&M University-Kingsville, John Howe Building,
Kingsville, TX 78363, USA}
\affil[3]{School of Natural Resources, University of Missouri, Anheuser-Busch Natural Resources Building
Columbia, Missouri 65211, USA}
\affil[*]{Email: tskb4@mail.missouri.edu, Phone: 847-274-1723}

\begin{document}

\maketitle

\textbf{Running Title}: Markov model for ACC behavior transitions

\newpage
\begin{abstract}
The use of accelerometers in wildlife tracking provides a fine-scale data source for understanding animal behavior and decision-making. Current methods in movement ecology focus on behavior as a driver of movement mechanisms. Our Markov model is a flexible and efficient method for inference related to effects on behavior that considers dependence between current and past behaviors. We applied this model to behavior data from six greater white-fronted geese (\textit{Anser albifrons frontalis}) during spring migration in mid-continent North America and considered likely drivers of behavior, including habitat, weather and time of day effects. We modeled the transitions between flying, feeding, stationary and walking behavior states using a first-order Bayesian Markov model. We introduced P{\'o}lya-Gamma latent variables for automatic sampling of the covariate coefficients from the posterior distribution and we calculated the odds ratios from the posterior samples. Our model provides a unifying framework for including both acceleration and Global Positioning System data. We found significant differences in behavioral transition rates among habitat types, diurnal behavior and behavioral changes due to weather. Our model provides straightforward inference of behavioral time allocation across used habitats, which is not amenable in activity budget or resource selection frameworks. 
\end{abstract}

\section{Key-words} 
animal behavior, auxiliary variables, hierarchical models, multinomial logistic, multiple imputation

\section{Introduction}

Animals make decisions daily that can result in differential fitness \citep[i.e., survival or reproductive success;][]{brown2004toward,breed2015animal}. The knowledge of an animal's behavior provides insight into its decision-making process. Historically, behavioral studies of animal populations were conducted using methods such as direct observation of focal individuals or instantaneous scan sampling of the group \citep{altmann1974observational}. However, data collection by direct observation prevents a comprehensive understanding of the decision-making process because it is limited to specific times and places when the animal is observable. The introduction of wildlife tracking devices has largely mitigated these constraints and provided observations over greater time and space. Recent improvements in tracking technology have allowed researchers to gather high frequency data over extended periods of time (e.g.,  >1 year), which has led to unprecedented insights into animal decision-making \citep{nathan2012using,leos2017analysis}. 

Behavioral inference from animal movement models has long been a research goal. Early work by \citet{morales2004extracting} showed the accuracy of movement state-space models based on location data only could be improved by including latent behavioral states allowing the model to switch between movement modes such as exploratory and encampment.  Specifically, an exploratory state was assumed to be characterized by longer step lengths and smaller turning angles, whereas an encamped state was characterized by shorter step lengths and larger turning angles. State-space models considering more than two states typically require more information. \citet{michelot2017estimation} included constraints in the temporal sequence of latent behavioral states based on expert knowledge of the annual cycle of the study species in order to distinguish among four states. \citet{mcclintock2017bridging} used a three state model, and states were distinguished by the addition of an auxiliary data stream. State estimation in this context typically is accomplished via hidden Markov models \citep{mcclintock2017incorporating}, which have proven to be an efficient method for estimating an unobserved sequence of a categorical variable, such as behavior, that is associated with the values of observed quantities \citep[e.g.,  step length and turning angle;][]{zucchini2016hidden}. In general, the estimated states are not interpretable as true behaviors, but rather proxies as they arise from a mixture model clustering procedure \citep{leos2017analysis,patterson2017statistical, michelot2019state}. Therefore, when ``ground truth'' data are available and inference on specific behavior states is the primary goal, behavioral inference is better achieved by building a classifier and then applying a behavior model to the classified states as we propose in this paper. 

The inclusion of accelerometers in wildlife tracking devices has become increasingly common. An accelerometer is a tool for measuring an object's acceleration (ACC), and when placed on animals, can be used to derive energy expenditure and behavior of the tagged individual independent of location information \citep{nathan2012using}. Data collected from accelerometers are substantially different from Global Positioning System (GPS) observations in quantity, resolution and quality. That is, ACC data can be collected at a high frequency throughout the life of the tracking device, which results in richer and relatively larger data sets. The frequency of ACC collection can range from nearly continuous to more widely spaced intervals and is typically set to the highest frequency possible before affecting battery performance. Importantly, unlike location tracking devices, accelerometers do not typically miss ``fixes'' because the accelerometer instrument does not require linkages with external equipment such as satellites or radio trackers. In addition, ACC data are collected in two or three axes of movement relative to the device position, which provides the ability to discriminate between behavioral states with similar trajectory profiles. 

The high frequency of data collected by accelerometers often allows researchers to ``ground-truth" acceleration profiles with visual observations of the animals. When ``ground-truth'' observations of behavior are available, behavioral classification of ACC data can be conducted using machine learning algorithms \citep{resheff2014accelerater, chakravarty2019novel}. These approaches have greater reliability at identifying more than two states when compared to methods that consider only location data \citep{resheff2014accelerater}. The richness of the ACC data allows for the derivation of many features to build the classification model. The fitted classification model can be used to assign behavior labels to ACC data. Then the labeled data can be used to make inference on animal behavior in a second stage. Currently, researchers summarize the labeled data within a time scale of interest (e.g.,  days or hours) and use a traditional activity budget framework, such as linking proportions of observed behaviors to covariates through generalized linear mixed models \citep{broekhuis2014optimal,heurich2014activity}. However, by analyzing the aggregate summarized data, inherent temporal structure and small scale processes in the data may not be fully utilized. The analysis of the activity budgets generally ignores temporal dependence in the data by modeling the proportions of behaviors separately. \citet{rugg1990analyzing} showed that a Markov model for behavior resulted in improved estimates of time allocation compared to traditional activity budget analyses. Additionally, the use of the labeled data ignores uncertainty associated with the chosen classification method. 


We propose a two-stage framework in which we first build a behavioral state classifier and then explicitly model temporal dependence in the high frequency behavioral observations via a Markov model and multiple imputation. Multiple imputation is often used in missing data scenarios and averages inference across a suite of potential true data sets \citep{scharf2017imputation,mcclintock2017incorporating}. We build a classifier using features of the accelerometer data then build potential data sets based on the classification probabilities or proportions. For a given behavior data set, our Markov model works directly on the behavioral state transitions and inherently assumes that current behavior depends on recent behavior (i.e.,  the Markov assumption states that given the most recent past, the current behavior is independent of the long-term past). The transition probabilities are modeled with a logistic function to link covariates to the probability of transition. Therefore, covariate effects retain the useful odds ratio interpretation as in multinomial logistic regression, and provide insight into how covariates affect the tradeoff between time spent in each behavior. By using additional habitat information from location data, we can infer habitat use differences among behaviors. Resource selection models determine selection of habitats by animals based on differences in frequency of use and availability. As compared to traditional resource selection frameworks, we can make these inferences without having to define an "availability distribution" \citep{hooten2017animal}. Although these approaches are important for identifying frequently used habitat, they do not define use and availability with respect to behavior. Our Markov model framework allows one to consider behavior profiles for different used habitats. Our approach provides the ability to answer questions about differential use of selected habitats that may be apparent from the focal animal's behavior. For instance, a resource selection study may identify two different agricultural crops as preferentially used, but these results do not identify the behavior associated with the crops or the differences in behavioral rates between the crops. 

The relationship between energy expenditure and fitness is especially critical for migratory animals because the time allocation of behaviors during migration, such as feeding, can impact survival during migration and subsequent reproductive success \citep{harrison2011carry}. We consider a long-distance migrant bird, the greater white-fronted goose (\textit{Anser albifrons frontalis}), to demonstrate the inference capabilities of our Markov model. The migration route of white-fronted geese spans a wide range of habitats in mid-continent North America. It is known that geese use different habitat types for different behaviors (e.g.,  roosting in water and feeding in crops) and thus, we expect differences in behavior transition probabilities or rates depending on habitat \citep{krapu1995spring}. Our Markov model has the ability to provide inference on potential differences in behavior transition rates among habitats that serve a similar purpose. For example, feeding may primarily occur in agricultural fields, but rates of feeding may differ by crop type. Variability in rates of behavior transitions may also be related to environmental factors such as weather. Although activity budget analyses during winter have not found strong effects of weather \citep{ely1992time}, studies during spring have found important relationships between weather and the timing of migratory movements \citep{fox2003spring}, and previous work has demonstrated heterogeneity in the movement of the geese \citep{hooten2018animal}. We include weather variables comprising temperature and wind because we anticipate that these features can explain variation in behavior transitions during spring migration (e.g.,  favorable weather conditions increase rates of flight). 

We implemented our Markov model in a Bayesian framework. At the core of our Markov model is multinomial logistic regression. There is a rich literature on Bayesian estimation of logistic models because sampling from the posterior distribution requires Metropolis-Hastings Markov Chain Monte Carlo algorithms that can be difficult to implement (tune) for moderate numbers of covariates or low observed frequencies of categories \citep{albert1993bayesian, holmes2006bayesian, fruhwirth2010data, polson2013bayesian}. To mitigate this challenge, automatic sampling strategies have been developed that rely on augmentation of data with latent variables. For example, \citet{albert1993bayesian} introduced the use of truncated normal random variables for category membership using probit link regression. \cite{hooten2010agent} used the Albert and Chib sampling for a multinomial movement model. \citet{holmes2006bayesian} expressed a multinomial logistic model as a product of binary models and used truncated scale mixture normals to define category membership. Alternatively, \citet{polson2013bayesian} developed a class of P{\'o}lya-Gamma distributions for automatic sampling of Bayesian logistic regression models that is faster and provides exact inference when compared to the scale mixture latent variables.  The introduction of P{\'o}lya-Gamma latent variables induces simple Gibbs updating steps for the logistic regression coefficients when using a normal (Gaussian) prior. Additionally, \citet{polson2013bayesian} showed the P{\'o}lya-Gamma scheme is significantly faster than other data augmentation methods and more efficient than Metropolis Hastings for mixed effects logistic models. Note that P{\'o}lya-Gamma latent variables have been used to estimate coefficients of a transition probability matrix of a hidden Markov model for rainfall data \citep{holsclaw2017bayesian} and to learn dependence structure in topic models for text mining applications \citep{chen2013scalable, linderman2015dependent, glynn2019bayesian}.  P{\'o}lya-Gamma latent variables have yet to be used for analysis of animal behavior data. Moreover, to the best of our knowledge, this is the first comprehensive example of P{\'o}lya-Gamma latent variables and multiple imputation within a Markov model.


The presented framework provides rich inference about covariate effects including information related to the location from GPS data on the sequence of observed behaviors derived from ACC data. We expected to estimate behavior transition probabilities for greater white-fronted geese during spring migration that confirm prior knowledge of the diurnal pattern of behavior and habitat use. The coefficients should reflect an increase in stationary behavior overnight in open water and wetland habitats while also suggesting more movement during the day. Additionally, we hypothesized that the estimates would exhibit variability among habitat types, suggesting variability in time allocation. Furthermore, we expected behavioral transitions associated with flight and feeding to be influenced by inclement weather more than walking and stationary behaviors. Our method provides a fine-scale picture of the behavioral decision-making process that is driven by both ACC and GPS data rather than mechanistic drivers of movement.

\section{Materials and Methods}
\subsection{Dataset}

\begin{figure}
	\centering
	\includegraphics[scale=0.75]{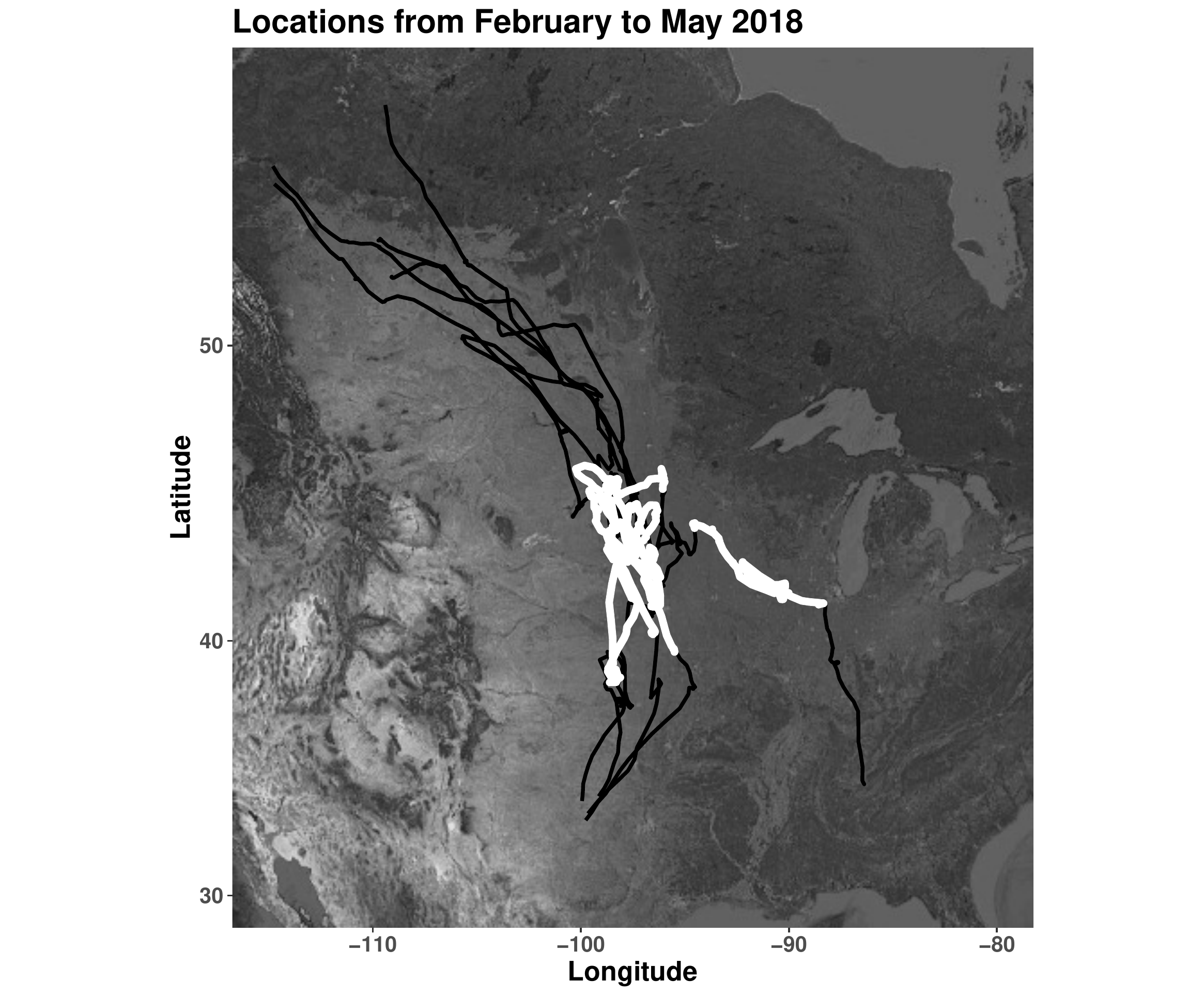}
	\caption{Spring migration paths in North America of 6 individual greater white-fronted geese (\textit{Anser albifrons frontalis}) from GPS equipped tracking devices. The highlighted section (white) are the locations for March 2018 used in the analysis of behavior transition probabilities.}
	\label{fig:map}
\end{figure}

Greater white-fronted geese, hereafter white-fronted geese, migrate from wintering areas in the southern US (i.e.,  Arkansas, Louisiana, Mississippi and Texas) to breeding areas in Alaska and northern Canada (Fig.\ref{fig:map}; \citealp{baldassarre2014ducks}). Individual female white-fronted geese were captured using rocket nets between December 2017 and January 2018 in Texas. Captured individuals were each fitted with a U.S. Geological Survey metal leg band and solar-powered Ornitela (\href{http://www.ornitela.com}{http://www.ornitela.com}) neck collar, comprising GPS, ACC and Global System for Mobile communications (i.e.,  for daily data upload) technology. Age, sex and morphometric measurements (normal wing cord [cm], head, culmen, tarsus, middle toe lengths [mm] and mass [nearest 0.1 kg]) were also collected. We set tracking devices to obtain ACC values at 10 Hz for 3 seconds at 6 minute intervals and GPS locations at 30 minute intervals with data upload every 24 hours. To demonstrate our modeling approach, we used GPS and ACC data from six white-fronted geese subset to 1-31 March 2018, which comprises a portion of the spring migration period (Fig.\ref{fig:map}). For simplicity, we chose a time period when all of the geese were migrating within the U.S. rather than defining the specific dates associated with the geese leaving wintering habitat and arriving at breeding areas. The average number of ACC fixes per individual was 7,257 with a range of 7,054 to 7,319. 

The location data were used to determine habitat and weather factors experienced by geese during spring migration. The U.S. Department of Agriculture maintains a raster of habitat and crop data throughout the contiguous lower 48 states (CropScape; \href{https://nassgeodata.gmu.edu/CropScape/}{https://nassgeodata.gmu.edu/CropScape/}), which is updated annually. We assumed the habitat at the GPS fix was best represented by the CropScape grid cell containing the observed point. The original CropScape categories were combined into fewer groups (supplementary materials S.1). The National Centers for Environmental Prediction's North American Regional Reanalysis (NARR; \href{https://rda.ucar.edu/datasets/ds608.0/}{https://rda.ucar.edu/datasets/ds608.0/}) data provide high resolution historical weather data in space and time. NARR is available eight times daily (i.e.,  every three hours) on an approximately 32 km grid. We assumed that the weather at the time of GPS fix was best represented by the NARR value corresponding to the nearest time and grid cell containing the observed point. We further assumed the weather and habitat at the time of an ACC fix were best represented by the values assigned to the most recent GPS location. The weather variables obtained were temperature [K], wind direction [ \degree ] and wind speed [ $m/s^2$ ]. Temperature was further summarized to daily minima and maxima. Weather variables were standardized to have mean 0 and standard deviation 1. We assumed diurnal variability in behavior patterns because geese move most predictably near dawn and dusk between roosting and feeding areas; thus, we included such variability by using the local solar time of GPS fixes. We captured the diurnal behavior of the geese through two continuous covariates with a 24 hour period calculated from the local solar time of day (in seconds) by $cos(2\pi (seconds)/86400)$ and $sin(2\pi (seconds)/86400)$, referred to as $\cos(\mbox{time})$ and $\sin(\mbox{time})$.  That is,  $\cos(\mbox{time})$ is a representation of night (high values) and day (low values), while $\sin(\mbox{time})$ represents the first half of the day (high values) and the second half of the day (low values). 

\subsubsection{Behavior Classification}

We summarized the ACC fixes into the 52 features described in the appendix of \citet{resheff2014accelerater}. The features were used to classify the ACC observations into four behavioral categories (i.e.,  flying, feeding, stationary and walking) using a random forest model. We chose a random forest model for its efficient handling of the large feature space. By using the random forest model, we assume the behavior labels are conditionally independent in time given the ACC data. The number of variables used to build the classification trees was varied from 1 to 15 and chosen by repeated 10-fold cross validation. The final model used 4 variables and had 96.5\% accuracy on the training data set acquired by video recording. The random forest model was used to predict classification probabilities for the ACC fixes from the time period of interest. From the classification probabilities, we constructed $M = 200$ possible data sets for multiple imputation. We discuss the details of imputation in the Model Fitting section. For more details on fitting the classification model, see supplementary materials (S.2). 

\subsection{Model}
\subsubsection{Behavior Transition Model}

Each imputation data set is a regularly spaced time series of categorical behaviors, $\mathcal{S}_{nT_n} = \{s_{n0},...,s_{nT_n}: s_{nt} \in \{1,...,J\}\}$, where $s_{nt}$ is the observed behavior category for individual $n$, $n = 1,...,N$, and time point $t$, $t = 0,1,...,T_n$, from the set of $J$ behavioral categories. A Markov model for categorical time series data is defined by a transition probability matrix, $\boldsymbol{P}_{nt}$, describing the time-varying transition probabilities between observed states $s_{n,t-1}$ and $s_{nt}$ as follows:
        \begin{equation}\label{eq:Pmat}
        \begin{gathered}
        \boldsymbol{P}_{nt}  \equiv 
        	\begin{bmatrix}
        p_{n11t} & p_{n12t} & \cdots & p_{n1Jt} \\
        	p_{n21t} & p_{n22t} & \cdots & p_{n2Jt} \\
             \vdots& \vdots  & \ddots & \vdots \\
        	p_{nJ1t} & p_{nJ2t} & \cdots & p_{nJJt} \\
        	\end{bmatrix},
        \end{gathered}
        \end{equation}
where $p_{nijt} \equiv P(s_{nt} = j | s_{n,t-1} = i)$, and the row probabilities sum to one, $\sum\limits_{j=1}^J p_{nijt} = 1$. Therefore, given the behavioral state at the previous time step ($t-1$), the current observed category (at time $t$) is modeled as a multinomial trial with probabilities from the corresponding row of the transition matrix \eqref{eq:Pmat}.  We let $y_{nijt}$ be an indicator for individual $n$'s  transition from state $i$ at time $t-1$ to state $j$ at time $t$, in other words, $y_{nijt} = 1(s_{nt} = j | s_{n,t-1} = i)$ is defined to be $1$ for the case when the $n^{th}$ individual is in state $i$ at time $t-1$ and state $j$ at time $t$, and $0$ otherwise.  Then the aggregated transition indices vector $\boldsymbol{y}_{nit} = 
	\begin{bmatrix}
	y_{ni1t}&
	\ldots &
	y_{niJt}
	\end{bmatrix}'$, along with the corresponding transition probabilities, $\boldsymbol{p}_{nit} =
	\begin{bmatrix}
	p_{ni1t} &
     \cdots  &
	p_{niJt}\\
	\end{bmatrix}'$, describe a multinomial trial.

Covariates are introduced by the multinomial logistic link function on the elements of the vector of transition probabilities, $\boldsymbol{p}_{nit}$, \citep[e.g.,  see][]{sung1,holsclaw2017bayesian}:
        \begin{equation}\label{eq:multlogit}
        \begin{gathered}
        \psi_{nijt} \equiv log\left((p_{nijt})/(p_{niJt})\right) = \textbf{x}'_{nt}\boldsymbol{\beta}_{ij}, \  \; \; \text{and} \ \\
         p_{nijt} = (exp(\psi_{nijt}))/(\sum_{k = 1}^{J}exp(\psi_{nikt})),\\
        \end{gathered}
        \end{equation}
where we assume the vector of parameter coefficients $\boldsymbol{\beta}_{iJ}=0$ for all $i = 1,2,...,J$ for identifiability. In our application, $J$ refers to the walking behavior. Therefore, coefficients for transitions to walking are 0 and do not appear in subsequent figures \ref{fig:habCIs} and \ref{fig:quant}. Note that covariates may vary by individual and/or time, which induces non-homogeneous transition probabilities. 

\subsubsection{Parameter Model}

The $B$-dimensional parameter vectors in the data model multinomial logistic functions, $\boldsymbol{\beta}_{ij}$, for $i = 1,2,...,J$ and $j = 1,2,...,J-1$  are partitioned into three components:
        \begin{equation}
        \boldsymbol{\beta}_{ij} = 
            	\begin{bmatrix}
            	(\boldsymbol{\alpha}_{ij})' &
		(\boldsymbol{\zeta}_{ij})' &
            	(\boldsymbol{\theta}_{ij})' \\
            	\end{bmatrix}'.
        \end{equation}
where $\boldsymbol{\alpha}_{ij}$ is an $(N-1)$-dimensional vector of random individual effects, $\boldsymbol{\zeta}_{ij}$ is an $H$-dimensional vector of habitat intercepts and $\boldsymbol{\theta}_{ij}$ are $B-(N-1+H)$-dimensional vectors of fixed quantitative covariate effects.
The random individual effects are subject to a ``sum to zero'' constraint and therefore we only sample $(N-1)$ coefficients. We assume independent normal distribution priors for each of the sampled coefficients as given below. The habitat coefficients for each transition are assumed to have the same mean, $\mu_{ij}$. This mean is equivalent to the average intercept and, therefore, each habitat coefficient can be interpreted as the mean plus the habitat effect, similar to a cell means model in ANOVA. The common mean is assigned a flat prior.  The hierarchical centering of these habitat parameters leads to less correlation between parameters compared to estimating $H$ habitat effects and an intercept in the regression model \citep{gilks1995markov}.  In addition, centering provides more interpretability compared to including a baseline intercept and $H -1$ coefficients.  In summary, the prior distributions for these coefficients are given by
        \begin{equation}\label{eq:prior}
        \begin{gathered}
            \boldsymbol{\alpha}_{ij} = 
            	\begin{bmatrix}
            	{\alpha}_{1 ij} &
            	\ldots &
            	{\alpha}_{(N-1)ij} \\
            	\end{bmatrix}' 
            	\sim N(\mathbf{0},\sigma_{\alpha_{ij}}^2 \mathbf{I}), \ \text{such that}  \sum_{n=1}^N \alpha_{nij} = 0,\\ 
	   \boldsymbol{\zeta}_{ij} = 
            	\begin{bmatrix}
            	{\zeta}_{1ij} &
            	\ldots &
            	{\zeta}_{Hij} \\
            	\end{bmatrix}' 
            	\sim N(\mu_{ij} \mathbf{1},\sigma_{\zeta_{ij}}^2 \mathbf{I}), \quad p(\mu_{ij}) \propto 1,\\ 
            \boldsymbol{\theta}_{ij} = 
            	\begin{bmatrix}
            	{\theta}_{1ij} &
            	\ldots &
            	{\theta}_{B-(N-1+H)ij} \\
            	\end{bmatrix}' 
            	\sim N(\mathbf{0},\sigma_{\theta_{ij}}^2 \mathbf{I}).
        \end{gathered}
	\end{equation}
where $\mathbf{0}$ and $\mathbf{1}$ are vectors of zeroes and ones, respectively.	Note, all prior variances ($\sigma_{\alpha_{ij}}^2, \sigma_{\zeta_{ij}}^2, \sigma_{\theta_{ij}}^2$) are fixed at 100 to induce a vague prior and reduce number of parameters needing to be estimated.

\subsection{Model Fitting}

The likelihood of each coefficient vector, $\boldsymbol{\beta}_{ij}$, is the product of multinomial logistic functions. As the dimension of $\boldsymbol{\beta}_{ij}$ increases, the tuning of a Metropolis-Hastings algorithm becomes increasingly difficult. Latent variable schemes provide ways to automatically sample from the posterior without tuning and are generalized from binomial logistic regression to multinomial logistic regression \citep{albert1993bayesian,holmes2006bayesian,fruhwirth2007auxiliary,polson2013bayesian}. Specifically, the latent variable schemes express the likelihood as a product of binary logistic functions by the following transformation:
        \begin{equation}\label{eq:likelihood}
        \begin{gathered}
            L(\boldsymbol{\beta}_{ij} \vert \boldsymbol{\beta}_{i,-j}, \mathcal{S}_1,...,\mathcal{S}_N) \propto \prod_{n=1}^N \prod_{t=1}^{T_n} \prod_{j=1}^{J-1} \left[ e^{\psi_{nijt} }/(\sum_{k = 1}^{J}e^{ \psi_{nikt} }) \right]^{y_{nijt} } ,\\
            \propto \prod_{n=1}^N \prod_{t=1}^{T_n} \left[ e^{\eta_{nijt} }/(1 + e^{ \eta_{nijt} }) \right]^{y_{nijt} } \left[1/(1 + e^{ \eta_{nijt} }) \right]^{1 - y_{nijt} },
        \end{gathered}
        \end{equation}
where $\eta_{nijt} = \psi_{nijt}  - C_{nijt}$ and $C_{nijt} = log\sum_{k \neq j} exp\{ \psi_{nikt}\}$.  In this formulation, $\eta_{nijt}$ is the log odds for a binomial random variable, which indicates whether or not the transition was from state $i$ to $j$ for individual $n$ at time point $t$ \citep{holmes2006bayesian,polson2013bayesian}.  From Theorem 1 of \citet{polson2013bayesian}, the product of logistic functions in the likelihood \eqref{eq:likelihood} is proportional to the product of of an exponential of $\eta_{nijt}$ and a P{\'o}lya-Gamma kernel. Therefore, by introducing the P{\'o}lya-Gamma latent variables, the sampling is done by conditional Gibbs updates alternating between the coefficients, $\boldsymbol{\beta}_{ij}$, and P{\'o}lya-Gamma latent variables. The details for the derivation of the full conditionals are in supplementary material (see supplementary materials S.3).


\subsubsection{Multiple Imputation}

The uncertainty associated with classifying the ACC data into categorical data by random forest is incorporated in the posterior distribution of the parameters via multiple imputation. Multiple imputation is used in situations involving missing data (e.g.,  the true behavior category) to provide approximate inference for the parameters based on the observed data (e.g.,  ACC fixes) \citep{rubin2004multiple, scharf2017imputation, mcclintock2017incorporating}. 

Let $\boldsymbol{\beta}$ be the collection of regression coefficients for all transitions, $\mathcal{S}$ be the collection of the state transitions for all individuals, and $\mathcal{A}$ be the collection of ACC data for all individuals. Our target distribution is the posterior of the unknown parameters, $\boldsymbol{\beta}$, given the behavior labels, $\mathcal{S}$. However, we do not observe the true behavior labels directly, but we predicted the behavior labels by classifying the observed auxiliary data,  $\mathcal{A}$. Therefore, our target distribution is

	\begin{equation}\label{eq:target}
	\begin{gathered}
		\left[\boldsymbol{\beta} \vert \mathcal{A} \right] = \int \left[\boldsymbol{\beta} \vert \mathcal{S}, \mathcal{A} \right] \left[\mathcal{S} \vert \mathcal{A} \right]d\mathcal{S}, \\
		= \int \left[\boldsymbol{\beta} \vert \mathcal{S}\right] \left[\mathcal{S} \vert \mathcal{A}\right] d\mathcal{S},
    	\end{gathered}	
	\end{equation}
where given the behavior labels, $\mathcal{S}$, the parameters are conditionally independent of the ACC data. We assumed the distribution of the behavior labels given the ACC data, $\left[\mathcal{S} \vert \mathcal{A}\right]$, is the prediction from the supervised classification random forest. 

We follow the multiple imputation MCMC algorithm outlined by \citet{scharf2017imputation} which numerically marginalizes over  $\left[\mathcal{S} \vert \mathcal{A}\right]$ by randomly sampling from a set of $M$ potential data sets where $M = 200$:
\begin{enumerate}
	\item Draw realizations from the imputation distribution, $\mathcal{S}^{(m)}_{1T_1},...,\mathcal{S}^{(m)}_{NT_N}$, for $m = 1,...,M$.
	\item For each iteration of the MCMC repeat:
	\halfblankline
	\begin{itemize}
		\item Select an imputation data set with probability $1/M$.
		\item Use the Gibbs updating steps in supplementary materials (S.3) with $(\mathcal{S}_{1T_1},...,\mathcal{S}_{NT_N}) = (\mathcal{S}^{(m)}_{1T_1},...,\mathcal{S}^{(m)}_{NT_N})$.
	\end{itemize}
\end{enumerate}

Throughout model fitting, we used ``walking'' as our reference behavioral category, $J$, in the multinomial logistic function \eqref{eq:multlogit}. The coefficients for the reference were set to $0$ and therefore are not displayed in the results. We chose walking as the reference category because we did not have specific hypotheses for this behavior and there was little variation in acceleration due to walking among individuals. Our method is amenable to choosing a different reference category specific to the study species. If it is of interest to more easily interpret effects on state duration, the reference category would depend on the previous state (i.e.,  for transitions from flight, the reference category would be flight).

We assessed parameter convergence by monitoring trace plots and setting different random starting values \citep{brooks1998general}. For inference, we sampled 15000 iterations from the model posterior and the first 5000 were discarded as burn in to ensure summaries were not influenced by the starting values. We generated behavior sequences from the current values of the parameters each iteration to investigate the goodness of fit of the model (supplementary web material S.4). 

The posteriors for odds ratios were obtained by exponentiating the coefficients each iteration. The posterior samples for coefficients and odds ratios were summarized to the posterior mean and 95\% credible intervals. We determined significance by whether the 95\% credible intervals included zero for the quantitative covariates for weather and time of day. The significance of the quantitative coefficients corresponded to the 95\% credible interval for the odds ratio not containing one and a proportion of >0.95 (positive effect on transition probability) or <0.05 (negative effect on transition probability) of iterations in which the odds ratio was >1. For each transition probability, we investigated the differences in behavioral transition probabilities among habitats by calculating the pairwise proportion of iterations in which habitat coefficients differed in magnitude, $\zeta_{aij} > \zeta_{bij}, \; a,b \in \{1,...,H\}, \; and \; a \neq b$. Two habitat coefficients for a transition were considered significantly different if the proportion of iterations with a difference in magnitude was >0.95 (the probability of transition from behavior $i$ to $j$ is greater in habitat $a$ than habitat $b$) or <0.05 (the probability of transition from behavior $i$ to $j$ is lesser in habitat $a$ than habitat $b$). We did not adjust the proportion cut off or widen credible intervals to account for inherent multiplicity in this case because we modeled the habitat coefficients with a common mean, which pooled the coefficient estimates \eqref{eq:prior}. The hierarchical centering shrinks estimates toward the common mean, $\mu_{ij}$, which makes it harder for significant pairwise differences to occur, thus eliminating the need to make additional post hoc adjustments for multiple comparisons \citep{gelman2012we}.

The assumption of a discrete time Markov process implicitly accommodates inference on the original sampling scale of the ACC data schedule (e.g.,  every 6 minutes). We interpreted all coefficient estimates as effects on behavioral transition probabilities from state $i$ to state $j$ at a 6 minute interval, relative to a base behavior state of walking. For brevity, we did not explicitly restate ``relative to walking at a 6 minute interval" for each interpretation. Similarly, the odds ratios are interpreted as the multiplicative change in odds of transitions to state $j$  from state $i$ versus transitions to walking from state $i$.

\section{Results}

\begin{figure}
	\centering
	\includegraphics[scale=0.65]{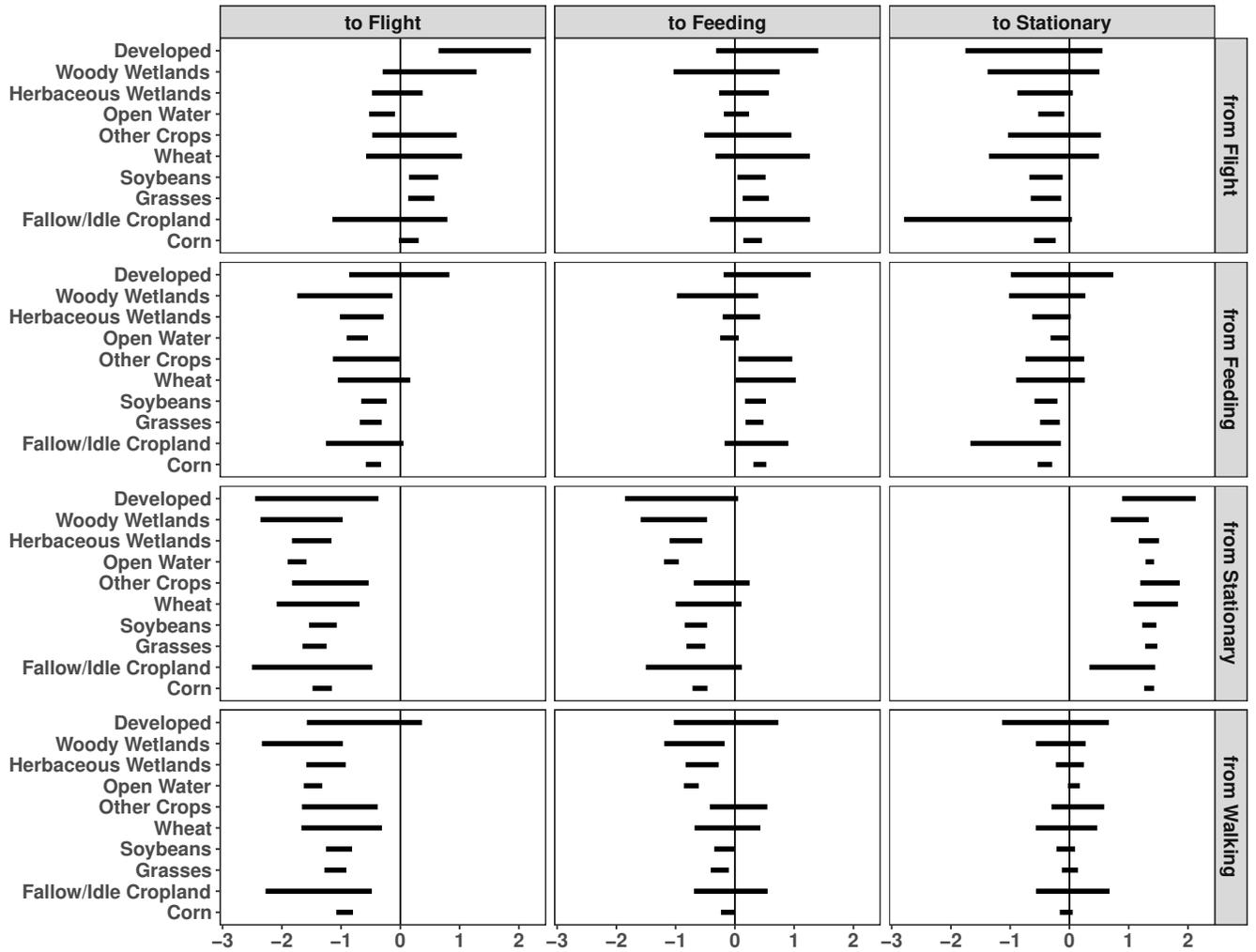}
	\caption{95\% credible intervals for habitat effects on the log-odds scale estimated in a Bayesian framework by behavior transition for six greater white-fronted geese in March 2018. There are significant pairwise differences between intervals that do not overlap.}
	\label{fig:habCIs}
\end{figure}

Our modeling framework establishes flexibility and efficiency in estimating covariate effects in the behavioral decision-making process by specifying transition-specific coefficients. The variability in coefficients across from states is indicative of the complexity in the decision-making process in migrating white-fronted geese. The estimated effects of different habitats on behavior transitions rates did not follow the same patterns across the from states (Fig. \ref{fig:habCIs}). By contrast, for weather and time of day covariates, the pattern in the coefficient estimates was similar across from states (Fig. \ref{fig:quant}). 

There were significant differences between habitats for every behavior transition probability except flight to stationary (see supplementary materials S.5). We estimated a significantly higher probability of transitioning to feeding from feeding (i.e., continued to feed), stationary and walking for food crops such as corn and soybeans and a significantly lower probability when the birds were in open water habitat (Fig. \ref{fig:propMat}). The positive coefficient estimates for corn and soybeans indicated an odds ratio >1 and an increase in the probability for transitioning to feeding from flight and feeding (Fig. \ref{fig:habCIs} to feeding subplots). For example, the mean effect of corn on the odds of continuing to feed was 0.42 with corresponding mean odds ratio 1.52, indicating the odds of continuing to feed at the next time step was 52\% greater than the odds of transitioning to walking from feeding in corn habitat (Table \ref{tab:habOR}). Transitions to stationary from feeding and walking were more probable in open water habitats than in corn (see supplementary materials S.5). In general, the widths of the credible intervals in Figure \ref{fig:habCIs} have an inverse relationship with the observed frequency of habitats (see supplementary materials S.1 for frequencies). For example, the most frequently used habitats were corn and open water, which tended to have the narrowest credible intervals. 

The probability of remaining in flight (i.e., transitioning to flight from flight) increased significantly during the first half of the day as indicated by the left most credible interval for sin(time) in the top left subplot of Fig. \ref{fig:diurnal}. There were few ACC fixes classified as flight at night, which corresponds to the significant negative effects of cos(time) on all transition probabilities to flight (Fig. \ref{fig:diurnal}). ACC fixes were classified as stationary most frequently and in greater proportions at night. Therefore, it is not surprising that many diurnal coefficients corresponding to transition probabilities to the stationary state were significant. Table \ref{tab:quantOR} shows that 6 out of 8 diurnal coefficients for transitions to stationary were significant. The transition probabilities to feeding and to flight from any behavior decreased overnight (negative effect of cos(time) across all from states to feeding and flight; Fig. \ref{fig:diurnal}).

Daily minimum and maximum temperatures did not affect the probability of remaining in flight the same way. Warmer than average daily minimum temperatures decreased the probability of remaining in flight (negative credible interval in top minimum temperature subplot of Fig. \ref{fig:weather}), while the warmer than average daily maximum temperatures increased the probability (positive credible interval in top maximum temperature subplot of Fig. \ref{fig:weather}). Increased wind speeds decreased the probability of remaining in flight and transitioning to feeding from feeding and walking. We found no evidence of the weather variables effecting transition probabilities from stationary behavior. The probability of remaining in flight was significantly affected by minimum daily temperature, maximum daily temperature and wind speed, but not wind direction (Table \ref{tab:quantOR}). 

\section{Discussion}

We provide a unified framework to connect variation in animal behavior with variation in habitat use and weather by propagating uncertainty in transitions using multiple imputation within a Bayesian Markov model with data from a long-distance migratory bird. Our approach is broadly applicable to other focal species and study systems across ecology. By analyzing data on the scale of frequency of collection, inferences are more intuitive and appropriate than aggregating to proportions. Importantly, our approach allows analysis of the behavior sequence with inherent temporal dependence and inference about covariate effects on behavior transition probabilities. 

Transition matrices are well studied and can provide a wealth of inference and prediction beyond what is presented in this study including simulation of behavior sequences in different settings. Further, transition matrix models are used in ecological and evolutionary research beyond behavioral applications and Markov models. We extended our Markov model with a P{\'o}lya-Gamma sampling scheme, which will be useful for fast and automatic estimation of other complex ecological models that utilize the logistic link function. 

We also implemented a unique approach to analyzing both ACC and GPS data from tracking devices. When we know the specific location of animals on the landscape, quantifying the effects of habitat on behavior transitions provides unprecedented information regarding the differential rates of behavior in specific habitats. We made simple assumptions when we constructed our covariates by assigning values to ACC fixes from the most recent GPS fix, but there is the potential to link to existing animal movement models for interpolation of location data at times of ACC observations. \citet{mcclintock2017bridging} used a continuous time correlated random walk model to predict locations at a regular time interval \citep{johnson2008continuous}. The locations could be predicted to the time points of the ACC fixes and covariate values could reflect the prediction location or be imputed from the prediction distribution \citep{hooten2010agent, scharf2017imputation}.

In the case of white-fronted geese during spring migration, different rates of behavior can be attributed to different habitats. Larger relative effects of habitats associated with food sources such as corn and soybeans on transition probabilities to feeding compared to effects of wetland habitats and open water aligns with previous knowledge of white-fronted goose ecology \citep{ely1992time, krapu1995spring}. Open water consistently had lower transition probabilities than food habitats for transitions to feeding and flight. Our finding that white-fronted geese were less likely to transition to feeding or flight in open water compared to food habitats was consistent with our expectations. Also, we found significant effects of weather on behavior transition probabilities of white-fronted geese during spring migration. Higher winds decreased the odds of remaining feeding or transitioning to feeding from stationary or walking behavior. Duration of flight behavior (flight to flight transitions) was the most influenced by weather. The opposite effects of minimum and maximum daily temperature on flight durations may be indicative of more complex decision-making processes by white-fronted geese. An increase in flight duration with an increase in maximum daily temperature aligns with our \textit{a priori} assumptions that these birds do not often migrate beyond the snow line during spring because food is relatively inaccessible under snow. 

\begin{figure}

	\begin{subfigure}[c]{0.5\textwidth}
	\centering
	\caption{}
        \includegraphics[width=\textwidth]{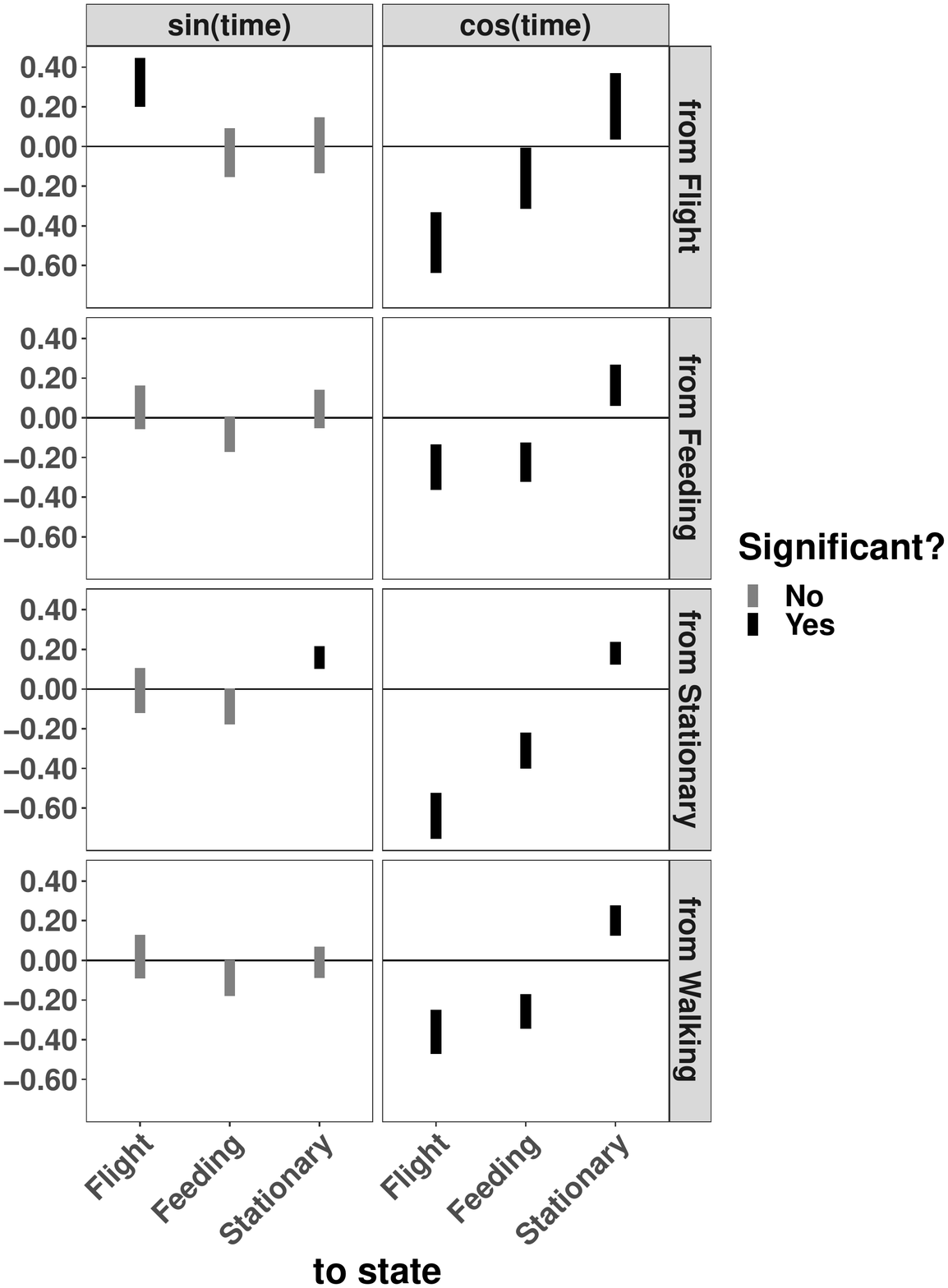}
        	\label{fig:diurnal}
    \end{subfigure}
        ~ 
    \begin{subfigure}[c]{0.5\textwidth}
    	\centering
    	\caption{}
        \includegraphics[width=\textwidth]{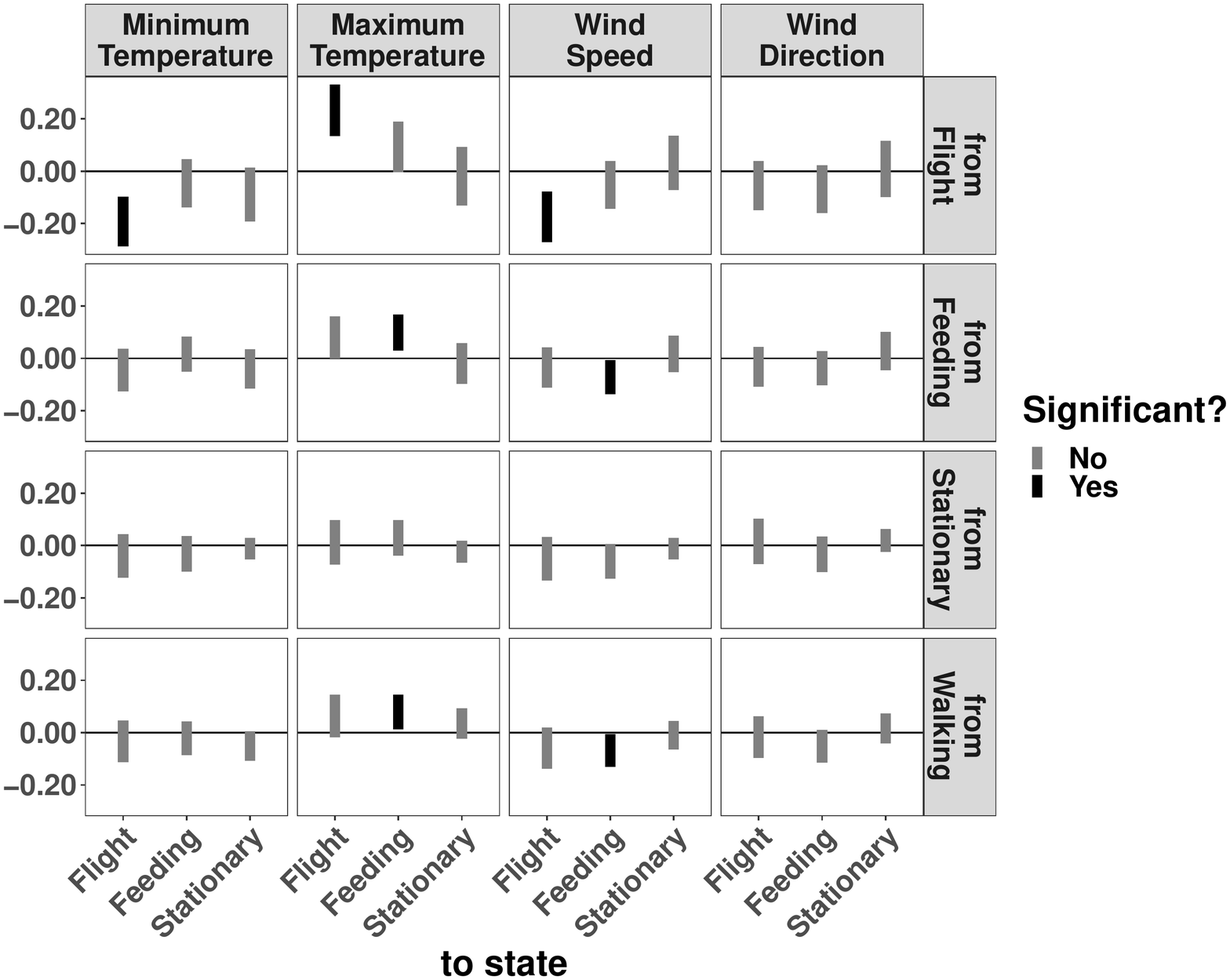}
        \label{fig:weather}
    \end{subfigure}
    	\caption{95\% credible intervals for diurnal (a) and weather (b) covariate effects on the log-odds scale estimated in a Bayesian framework by behavior transition for six greater white-fronted geese in March 2018. Significance refers to whether or not the credible interval overlaps zero.}
	\label{fig:quant}
\end{figure}

\begin{figure}
	\centering
	\includegraphics[scale=0.65,page=5]{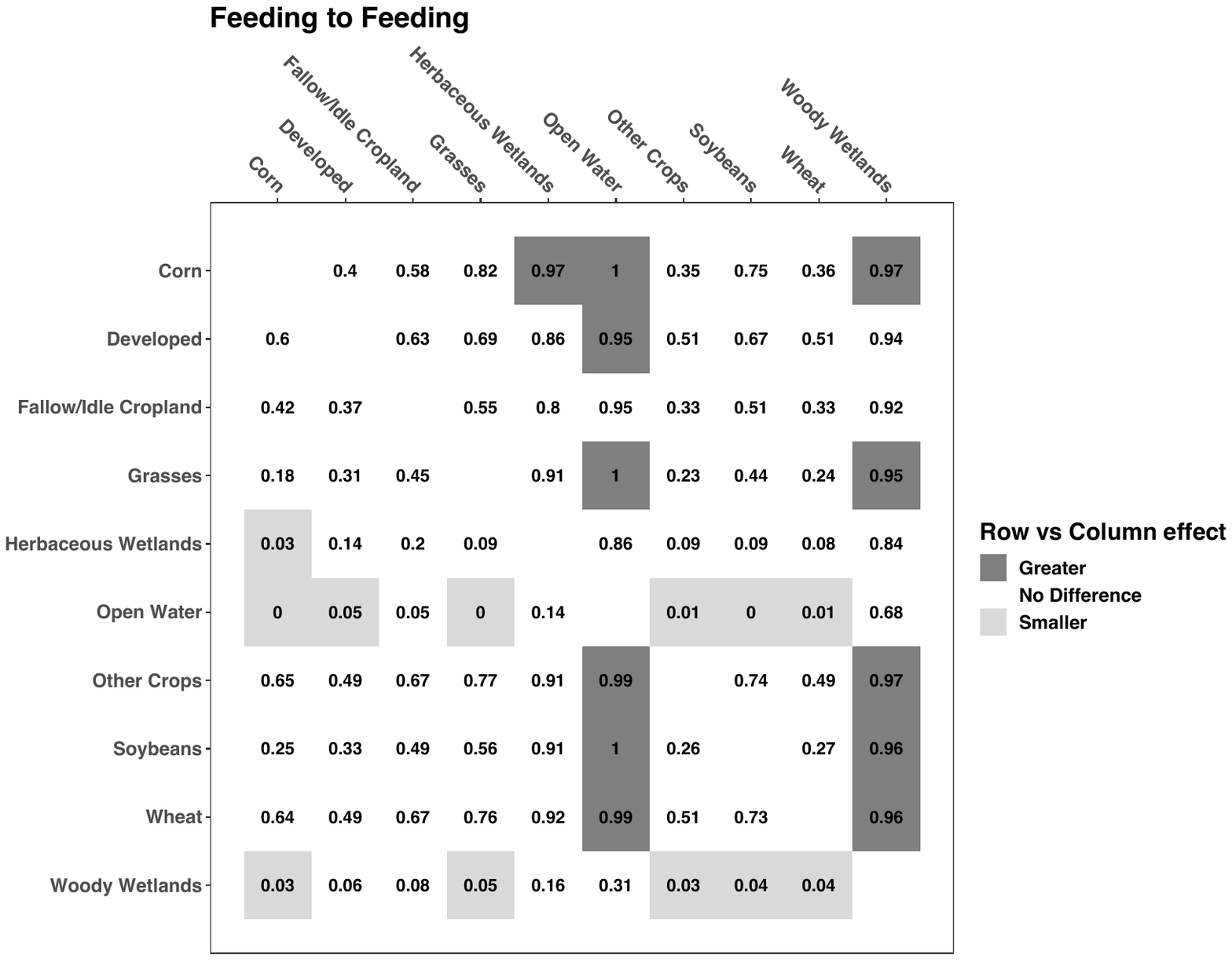}
	\caption{Matrix of pairwise comparisons between habitat coefficients for the transition feeding to feeding for six greater white-fronted geese in March 2018. The values indicate the proportion of posterior MCMC iterations in which the habitat coefficient down the row was greater than the coefficients along the column; the upper triangular values and lower triangular values sum to 1. For example, in the first row, corn, the proportion of times the estimate for the corn habitat coefficient was greater than the open water habitat coefficient was 1. The rest of the transitions can be found in the supplementary web material (S.5)}
	\label{fig:propMat}
\end{figure}

Furthermore, we were able to control for diurnal patterns in activity by using a continuous transformation rather than discretizing the time of day. The effects of time of day were consistent across from states. Specifically, the coefficients for cos(time) all appeared to be significant in the same direction and magnitude in Figure \ref{fig:diurnal}. In the future, it may be beneficial to relax the assumption that all covariate effects are transition specific and instead estimate a mix of transition-specific and behavior-specific coefficients. For example, \citet{holsclaw2017bayesian} estimated a transition matrix for a Hidden Markov model with a transition-specific intercept and state specific coefficients for all covariates. They estimated weather covariates specific to rainfall states at weather stations in India to identify global effects of weather systems. For animal migration and behavior, the occurrence of certain weather systems or the time of day may always result in a specific decision regardless of the current behavior which suggests a model with behavior specific coefficients may be a better fit. Within the Bayesian framework, models with different parameter formulations can be compared using tools such as Bayes factors, information criteria, or prediction of hold out samples to test hypotheses about the behavior process \citep{hooten2015guide}. 

Our analysis was limited to a month-long subset of the spring 2018 migration and this pattern may become more clear with the inclusion of more years. Although much of the inference was verification of previous knowledge about white-fronted geese, the methodology allows us to infer about habitat use and behavior simultaneously using both the GPS and ACC information. Most importantly, we developed a detailed picture of time allocation in reference to the specific habitat types used by white-fronted geese during spring migration which is not addressed by traditional models used in previous activity budget analysis, movement trajectory prediction, or resource selection frameworks.

There is diminishing return on estimation and inference after a sufficiently large number of imputation data sets are used \citep{scharf2017imputation, mcclintock2017incorporating}. In order to investigate our choice of 200 data sets, we compared posterior inference among 200 imputation data sets, 100 imputation data sets and 1 data set corresponding to the most likely behavior classification. There was a general consensus among the three scenarios on the direction of the effects. Compared to the estimates based on the most likely classification, the estimates for the imputation scenarios were shrunk towards zero. Credible interval widths for the imputation scenarios appeared consistent or larger than the widths for coefficients estimated from the most likely behaviors (see supplementary web material (S.6) for side-by-side comparisons). 

Advances in animal tracking technologies continue to provide more frequently collected data for a greater duration of time. Thus, rich data sets are emerging as never before for ecologists and evolutionary biologists. Hence, there is an increasing need for development of models that appropriately handle the structure and volume of collected information for improved inference. Our Markov model framework provides much more capability for directly interpreting behavior patterns. In addition, the P{\'o}lya-Gamma latent variables facilitate for more efficient sampling and have yet to be used in the animal behavior and movement literature. If classification of behaviors is not feasible, the P{\'o}lya-Gamma sampling scheme can be incorporated into Bayesian estimation of transition probability matrices in a hidden Markov model framework \citep{holsclaw2017bayesian}. The model directly handles temporal dependence in ACC data and learns about the behavior process from both ACC and GPS data. Our results suggest that new data sources coupled with appropriate modeling have unprecedented potential to provide a comprehensive understanding of complex ecological and evolutionary processes in animal movement.

\newpage 

\begin{table}[htp]
    \caption{Means and 95\% credible intervals for odds ratios for habitat coefficients for the feeding to feeding transition arranged by increasing mean estimated in a Bayesian framework for six greater white-fronted geese in March 2018. A significant difference between habitats is indicated by non-overlapping credible intervals which can be visualized in Fig. \ref{fig:propMat}. The values correspond to exponentiating the estimates of the coefficients depicted in the from feeding to feeding panel of Fig. \ref{fig:habCIs}}
    \begin{center}
    \begin{tabular}{cccc}
    \toprule
     Covariate & Mean & \multicolumn{2}{c}{95\% CI}\\
    \midrule
      Woody Wetlands & 0.80 & 0.38 & 1.48\\
    
      Open Water & 0.91 & 0.78 & 1.07\\
    
      Herbaceous Wetlands & 1.13 & 0.81 & 1.53\\
    
     Grasses & 1.40 & 1.20 & 1.62\\
    
      Soybeans & 1.42 & 1.19 & 1.69\\
    
      Fallow/Idle Cropland & 1.49 & 0.84 & 2.46\\
    
      Corn & 1.52 & 1.37 & 1.70\\
    
      Other Crops & 1.72 & 1.06 & 2.64\\
    
      Wheat & 1.75 & 1.01 & 2.79\\
    
    Developed & 1.82 & 0.83 & 3.59\\
    \bottomrule
    \end{tabular}

        \end{center}
\label{tab:habOR}
\end{table}

\begin{table}[htp]
    \caption{A selection of means, 95\% credible intervals and proportion of samples with an estimate greater than 1 for the odds ratios for quantitative covariate coefficients estimated in a Bayesian framework for six greater white-fronted geese in March 2018. The quantitative variable has a significant effect on the transition probability if the credible interval does not overlap with 1 which corresponds to a credible interval in Fig. \ref{fig:quant} not overlapping 0.}
    \begin{center}
    
            \begin{tabular}{cccccc}
            \toprule
            Transition & Covariate & mean &  \multicolumn{2}{c}{95\% CI} & Proportion of Samples > 1\\
            \midrule
             & Maximum Temperature & 1.26 & 1.14 & 1.39 & 1.00\\
             
             & Minimum Temperature & 0.83 & 0.75 & 0.91 & 0.00\\
            
            \multirow{-3}{*}{\centering\arraybackslash Flight to Flight}  & Wind Speed & 0.84 & 0.76 & 0.92 & 0.00\\
            
            \cmidrule{1-6}
            Flight to Stationary & & 1.23 & 1.04 & 1.45 & 0.99\\
            
            Feeding to Stationary &  & 1.18 & 1.06 & 1.31 & 1.00\\
            
            Stationary to Stationary &  & 1.20 & 1.13 & 1.27 & 1.00\\
            
            Walking to Stationary & \multirow{-4}{*}{\centering\arraybackslash cos(time)} & 1.22 & 1.13 & 1.32 & 1.00\\

            Stationary to Stationary & sin(time) & 1.17 & 1.11 & 1.24 & 1.00\\

%
%
%
            \bottomrule
            \end{tabular}
    
%
%
%
%
%
%
%
%

        \end{center}
\label{tab:quantOR}
\end{table}

%
%


\newpage

\appendix


\section*{Supplementary material}

# Supplemental Web Material: The utility of a Bayesian Markov model with Pólya-Gamma sampling for estimating individual behavior transition probabilities from accelerometer classifications

### S.1: CropScape Habitat Categories

We combined CropScape categories obtained from the U.S. Department of Agriculture (CropScape; https://nassgeodata.gmu.edu/CropScape/) into fewer categories in order to reduce the number of estimable habitat coefficients. Table 1 provides details on which categories were combined along with the frequency of each category associated with the ACC fixes.

Table 1: Habitat categories used in the Bayesian Markov model with the associated CropScape categories from the USDA. CropScape categories were matched to GPS data for six greater white-fronted geese during March 2018.

| Habitat Category | CropScape Categories | Frequency |
| --- | --- | --- |
| Corn | Corn, Pop or Orn Corn | 12852 |
| Developed | Developed\High Intensity, Developed\Med Intensity, Developed\Low Intensity, Developed\Open Space | 306 |
| Fallow\Idle Cropland | Fallow\Idle Cropland | 346 |
| Grasses | Alfalfa, Grass\Pasture, Other Hay\Non Alfalfa, Switchgrass | 7557 |
| Herbaceous Wetlands | Herbaceous Wetlands | 2242 |
| Open Water | Open Water | 12874 |
| Other Crops | Dry Beans, Millet, Oats, Peas, Rye, Sorghum, Sugarbeets, Sunflower | 760 |
| Soybeans | Soybeans | 5447 |
| Wheat | Spring Wheat, Winter Wheat | 571 |
| Woody Wetlands | Woody Wetlands | 589 |

## S.2: Acceleration Classification

The training data consisted of 481 raw acceleration fixes with "ground-truthed" behavior classifications of flight(150), feeding (106), stationary (150), and walking (75). Each raw acceleration ($m/s^2$) fix was a multivariate time series collected at 10Hz for 3 seconds (i.e. 30 observations of acceleration in the x, y, and z direction). The fixes were summarized into the 52 features described in the appendix of Resheff et al. (2014) by using the associated web tool (http://smell.huji.ac.il). Additionally, the web tool fits 8 models: KNN, linear SVM, RBF kernel SVM, decision tree, random forest, naïve Bayes, LDA, and QDA. From the default settings, random forest performed the best and was chosen as the final classification method. The final random forest model was tuned offline in R. The number of trees was set to 500 (i.e. the default in R package randomForest). The number of variables to try was varied from 1 to 15 and chosen by repeating 10-fold cross validation 3 times. The final model chosen used 4 variables and had 96.5% accuracy on the training data set acquired by video recording. The top five variables based on importance as measured by the mean Gini index are the wave amplitude of the acceleration in the y axis, overall dynamic body acceleration (ODBA), dynamic body acceleration (DBA) in the x axis, the $25^{th}$ percentile of the acceleration in the y axis, and the mean of the acceleration in the y axis (Figure 2).

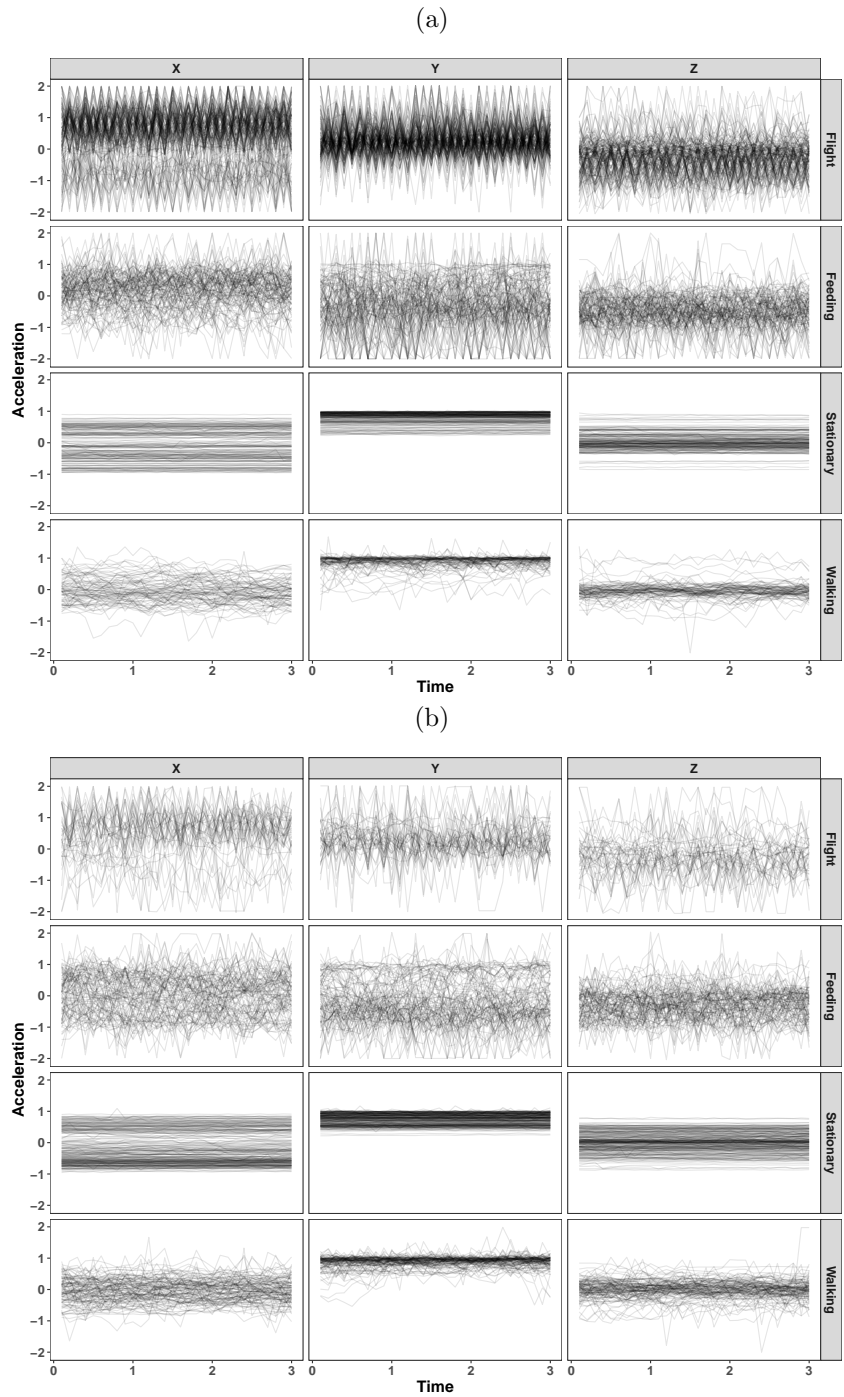

Figure 1: Raw acceleration time series for (a) the 481 training observations in the X, Y, and Z axis for the four behavior categories flight (150), feeding (106), stationary (150), and walking (75) and (b) a sample of 1% of the classified observations. In (b), the behaviors correspond to the most likely category based on the random forest predicted probabilities.

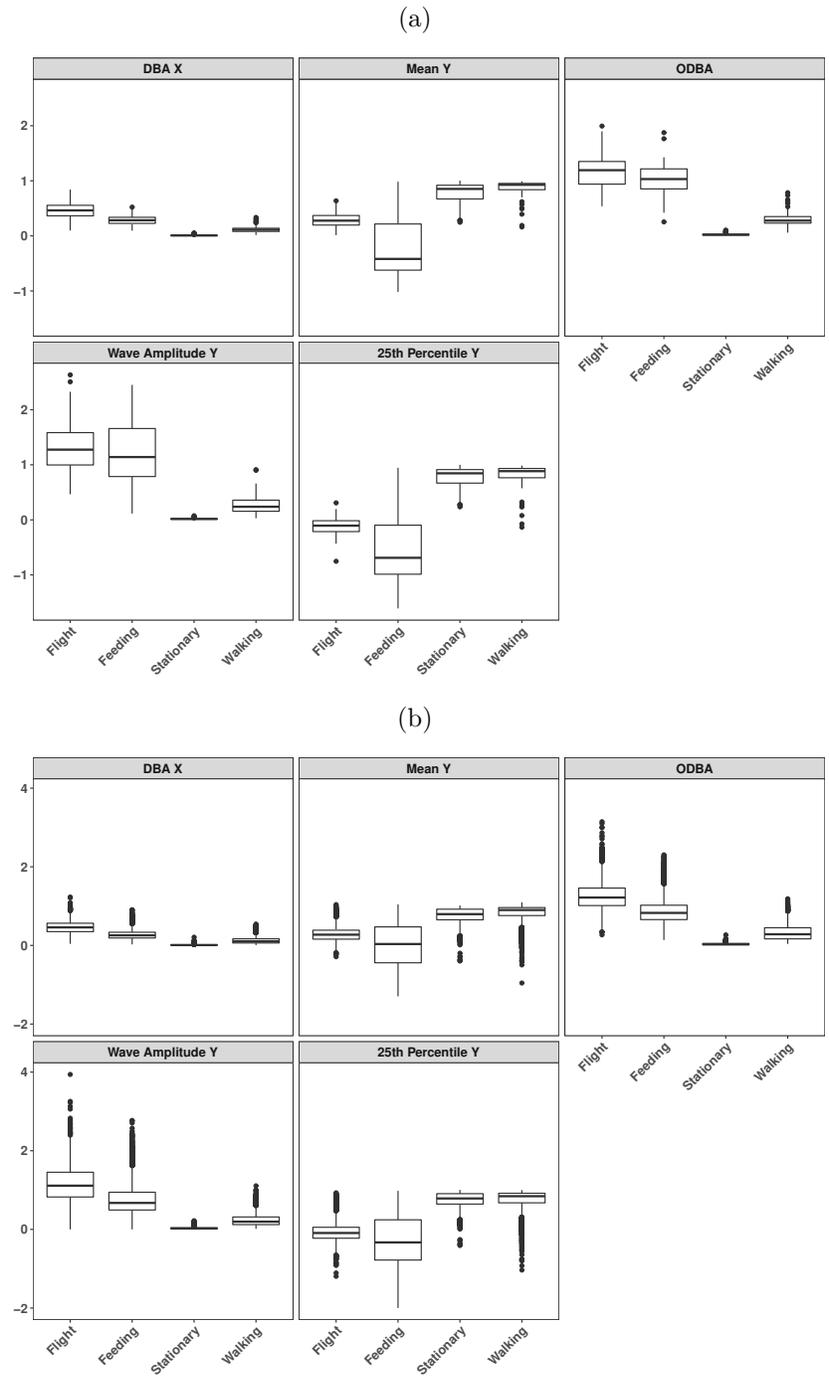

Figure 2: Box plots for the top five variables used in the classification random forest for the ground-truthed behaviors (a) and the classified observations (b). In (b), the behaviors correspond to the most likely category based on the random forest predicted probabilities.

## S.3: Conditional Distributions

Our model contains $J$ separate Bayesian multinomial logistic regressions with $T_i$, i = 1,...,J, observations where $T_i$ is the number of transitions from category $i$. The covariate matrix, $\boldsymbol{X}_i$, is $T_i \times B$, with rows $\mathbf{x}_{nt}$ and response vector $\boldsymbol{Y}_{ij}$ is a $T_i \times 1$ vector with elements $y_{nijt}$ where $t$ corresponds to a transition from state $i$. The Gibbs updating steps are then:

$$\begin{aligned}(\boldsymbol{\beta}_{ij}|\boldsymbol{\Omega}_{ij}) &\sim N(\mathbf{m}_{ij}, \mathbf{V}_{ij}), \\ (\omega_{nijt}|\boldsymbol{\beta}_{ij}) &\sim PG(1, \eta_{nijt}),\end{aligned} \quad (1)$$

where $\boldsymbol{V}_{ij}^{-1} = \mathbf{X}_i' \boldsymbol{\Omega}_{ij} \mathbf{X}_i + \mathbf{V}_0^{-1}$ and $\boldsymbol{m}_{ij} = \mathbf{V}_{ij}(\mathbf{X}_i'((\mathbf{Y}_{ij} - 1/2) + \boldsymbol{\Omega}_{ij}\boldsymbol{C}_{ij}) + \mathbf{V}_0^{-1}\mathbf{m}_0)$. Here, $\boldsymbol{\Omega}_{ij}$ is a $T_i \times T_i$ diagonal matrix with $\omega_{nijt}$ along the diagonal and $\boldsymbol{C}_{ij}$ is a $T_i \times 1$ vector with elements $C_{nijt}$ where t indexes transitions from category $i$. In this case, $\boldsymbol{V}_0$ and $\boldsymbol{m}_0$ correspond to the prior variance matrix and mean vector for $\boldsymbol{\beta}_{ij}$, respectively.

## S.4: Model Validation

We validated the model by investigating the posterior distribution of the frequency of each behavior transition. For each iteration from the posterior distribution, we calculated the transition probabilities using the current values of the parameters and the covariates. Using the transition probabilities, we generated one sequence of behaviors for each individual. Then, we summarized each generated sequence into the behavior transition frequencies to build the posterior distribution. We also calculated the frequency of each transition for the 200 imputation data sets to build an "imputed" distribution. As shown in the figures on the next page, the posterior distribution covers the imputation distribution for every transition and individual suggesting the model is a good fit to the data.

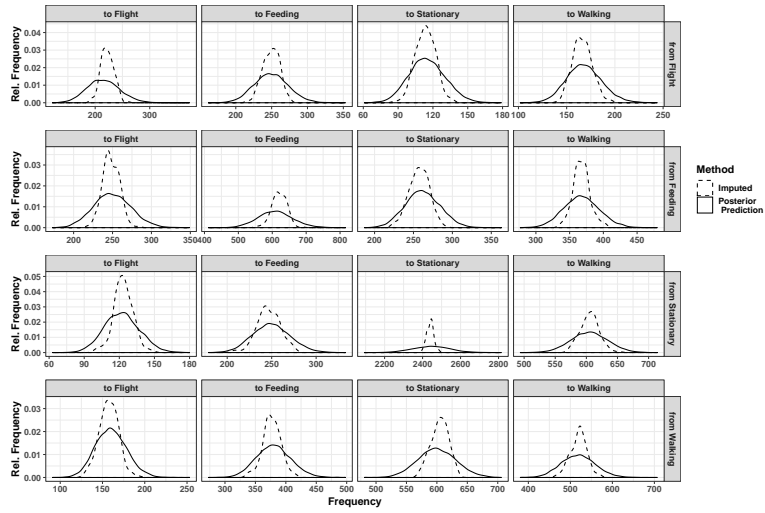
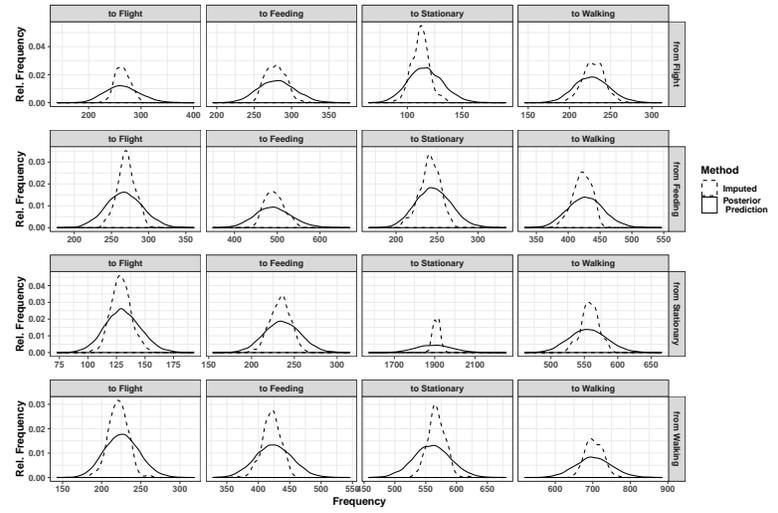
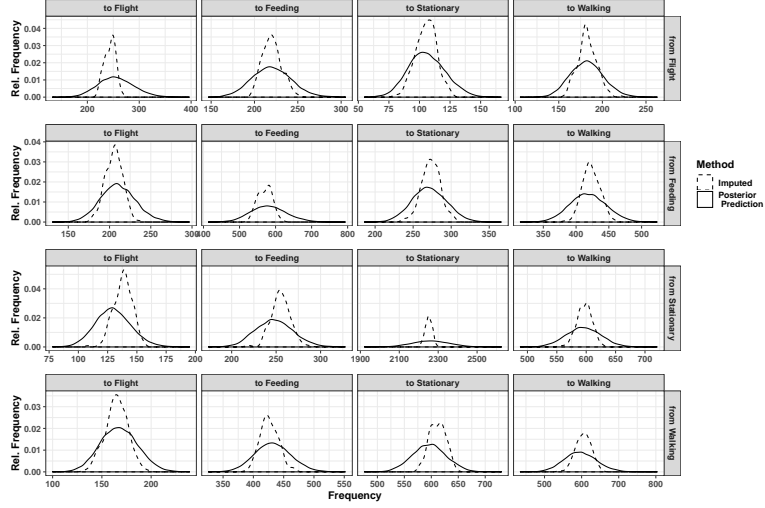
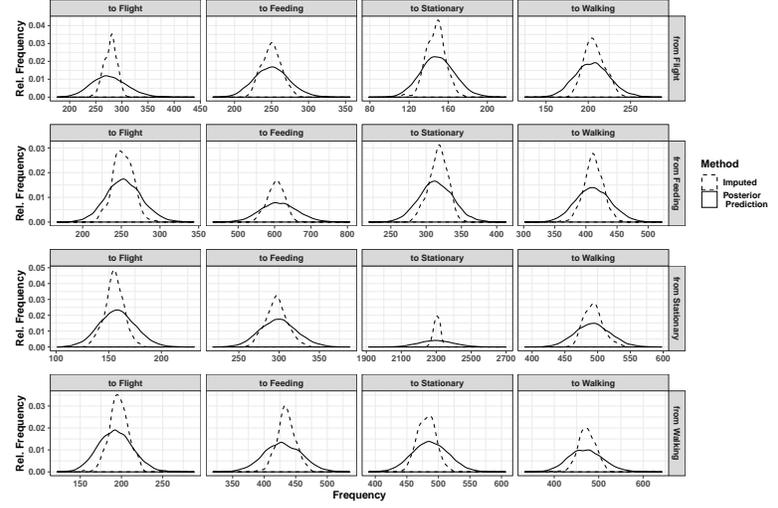
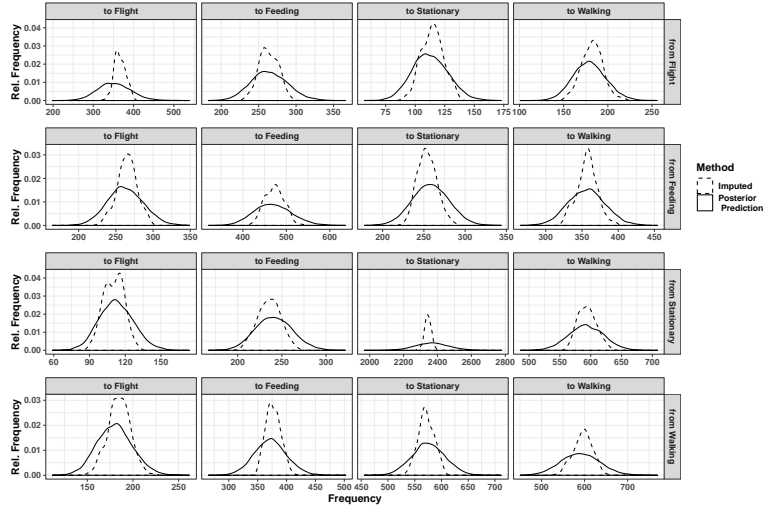
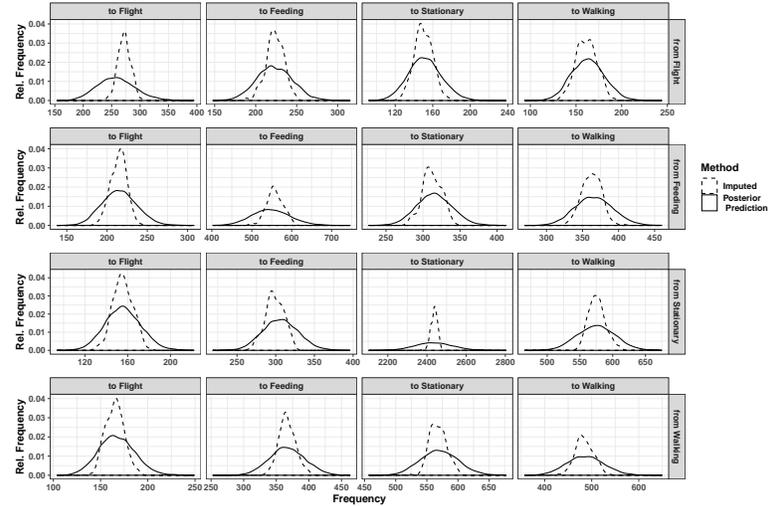

## S.5: Habitat Multiple Comparison Matrices

The following matrices of pairwise comparisons between habitat coefficients visualize the proportion of samples from the posterior distribution in which the habitat coefficients differed in magnitude. The values indicate the proportion of samples in which the habitat coefficient down the row was greater than the coefficients along the column; the upper triangular values and lower triangular values sum to 1. For example in the first matrix, in the first row, corn, the proportion of times the estimate for the corn habitat coefficient was greater than the open water habitat coefficient was 1. The transition from grazing to flight was the only transition to not have any significant pairwise differences among habitats.

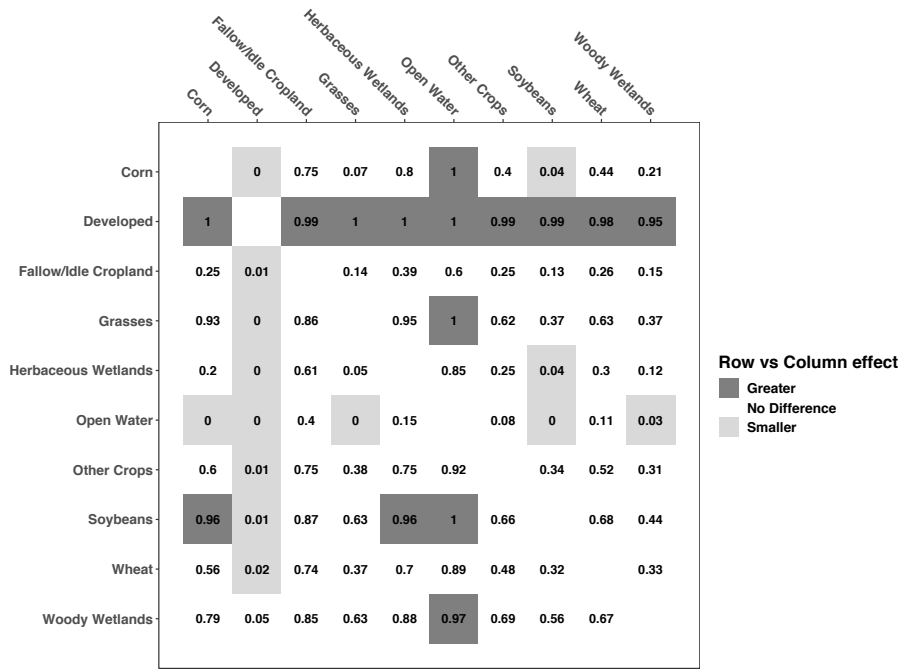
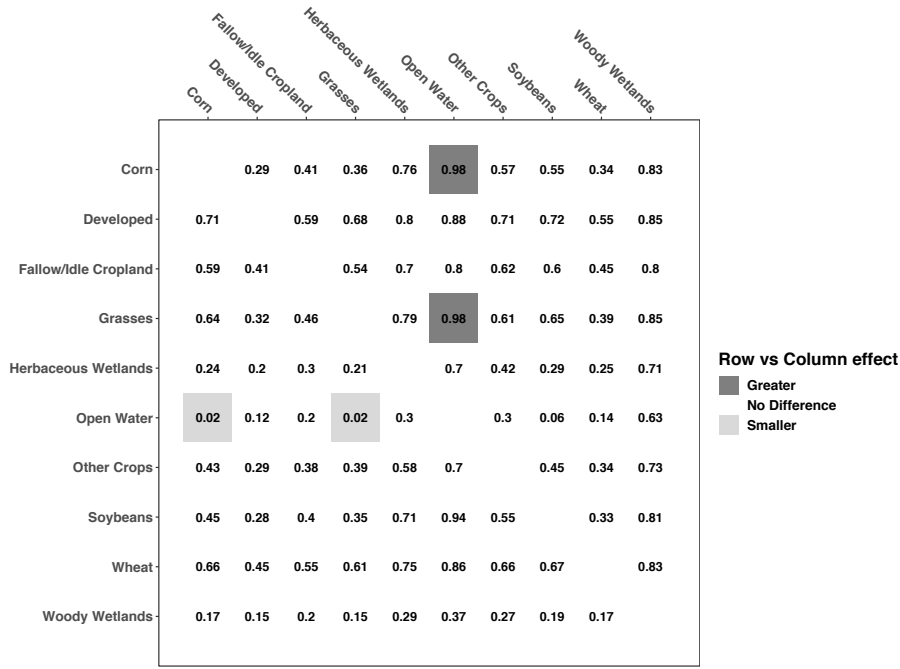
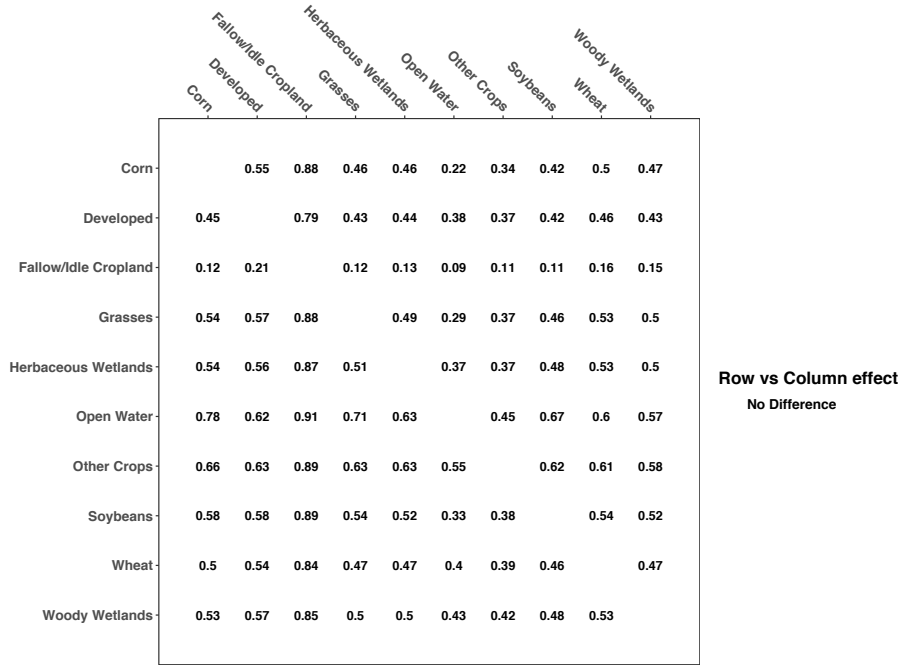

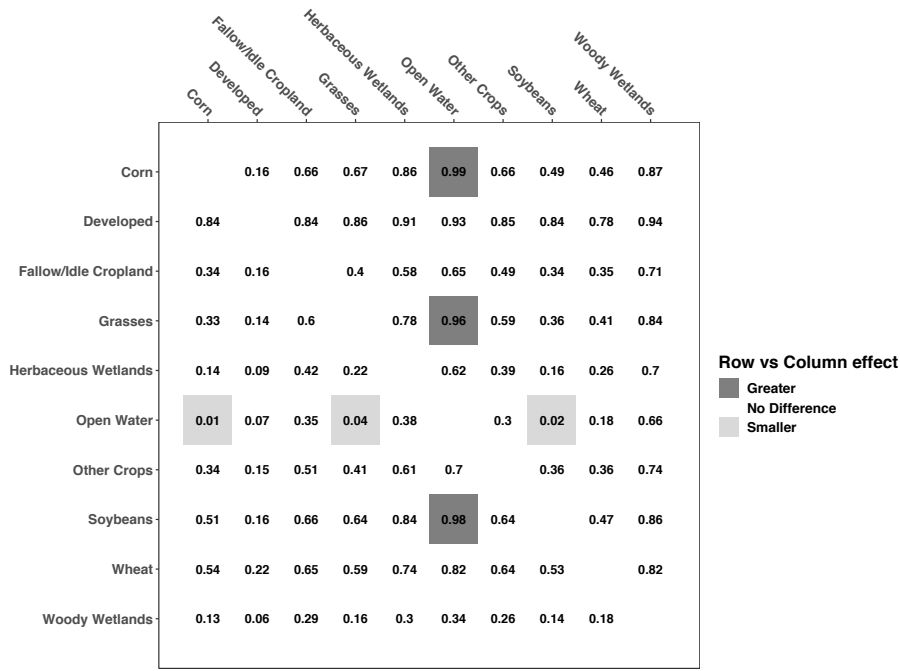

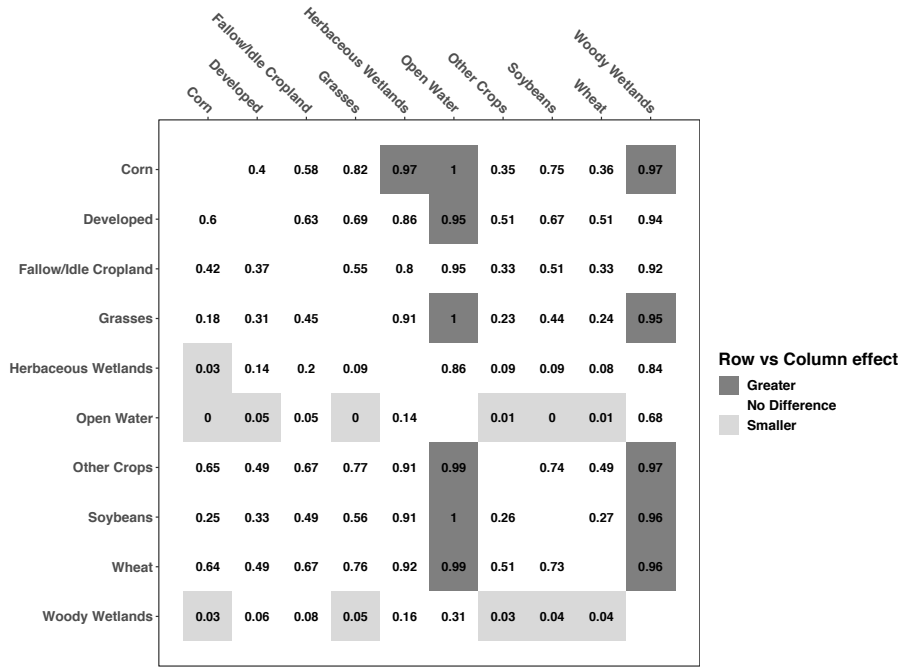

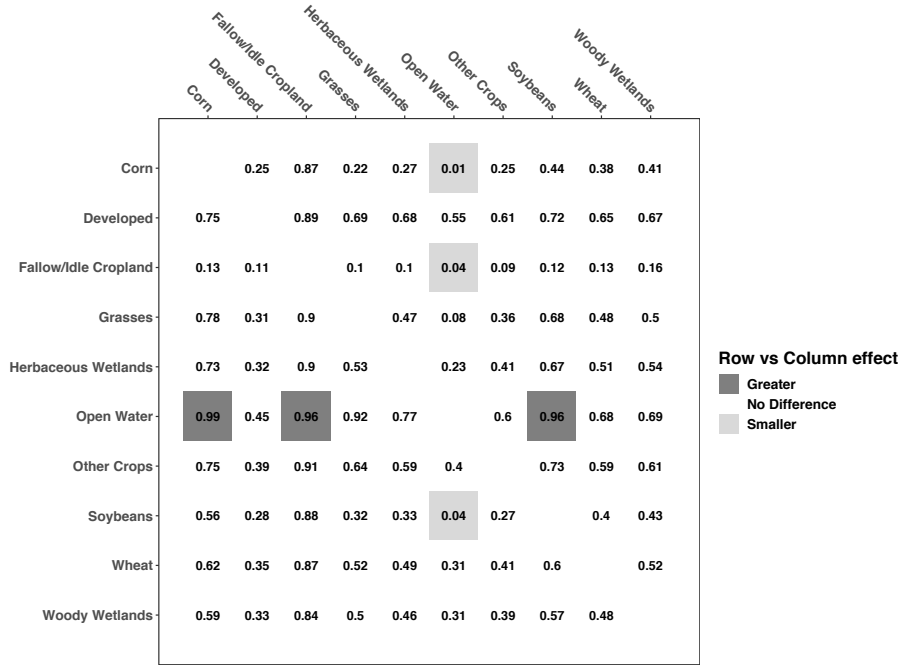

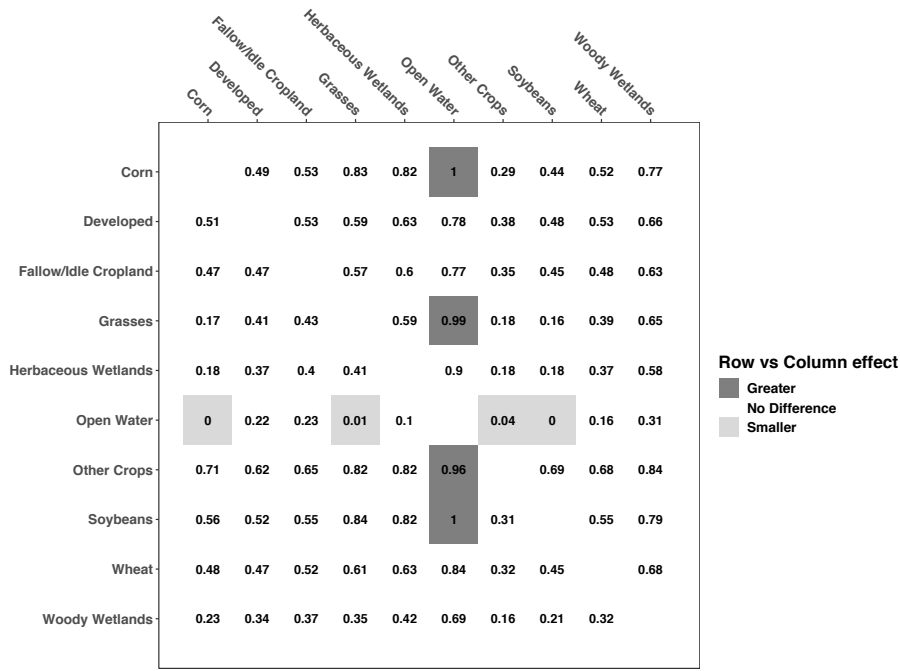
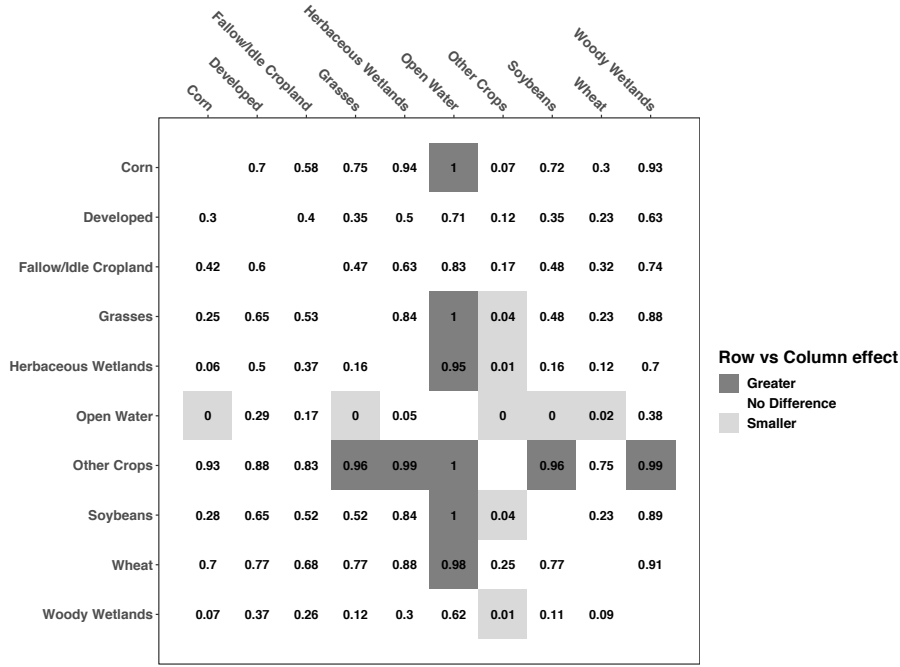
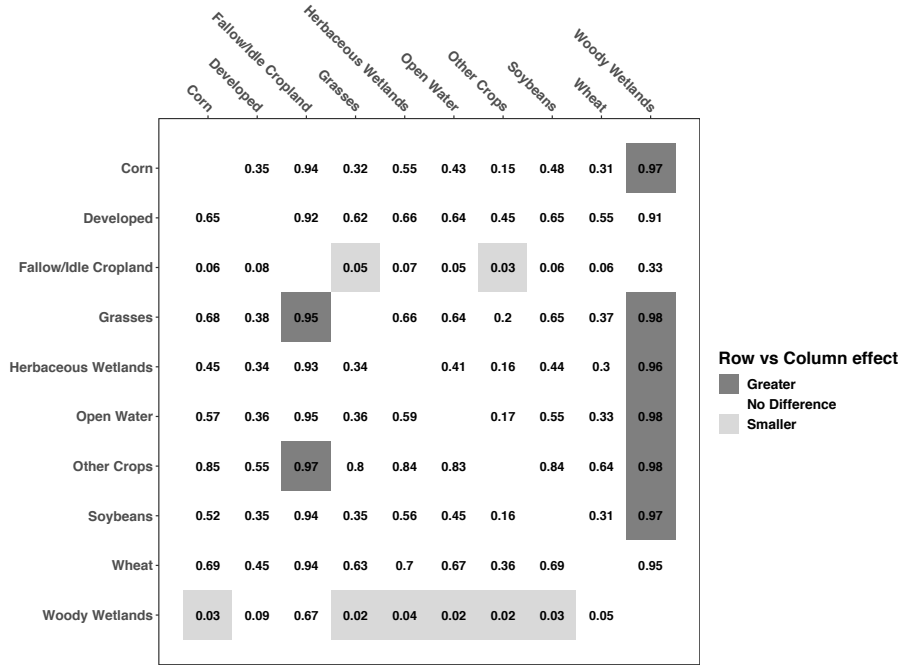

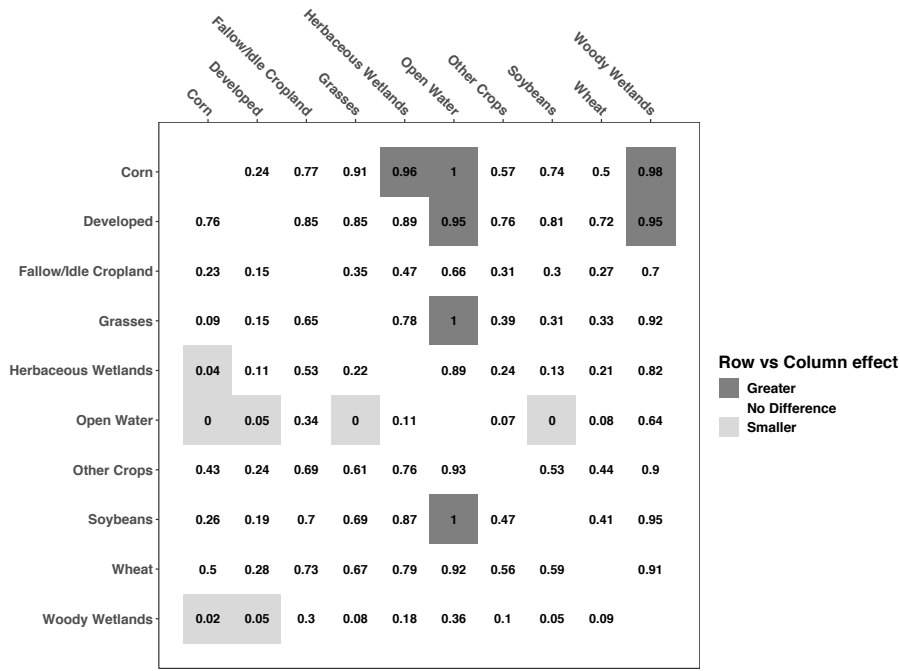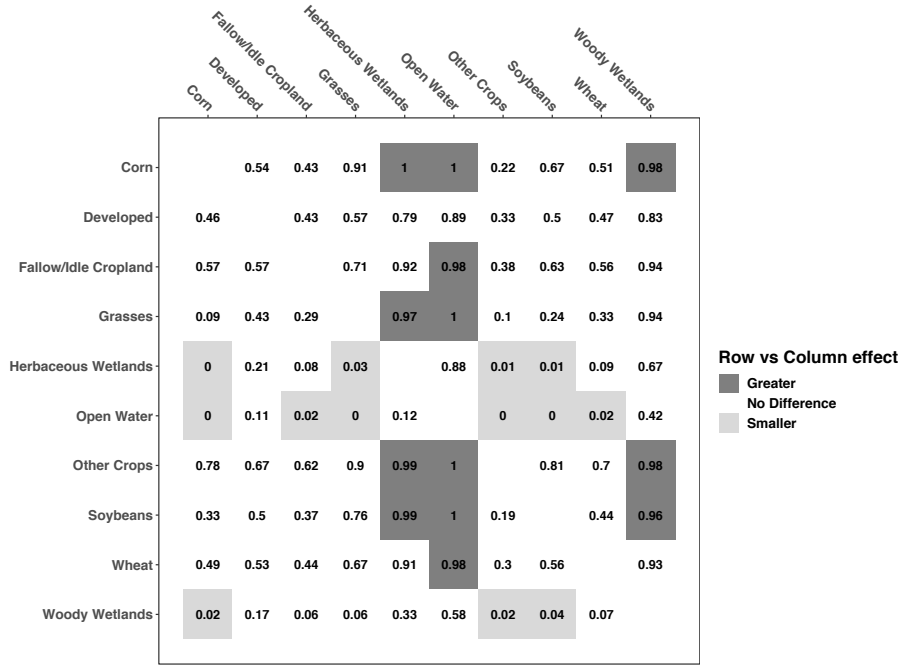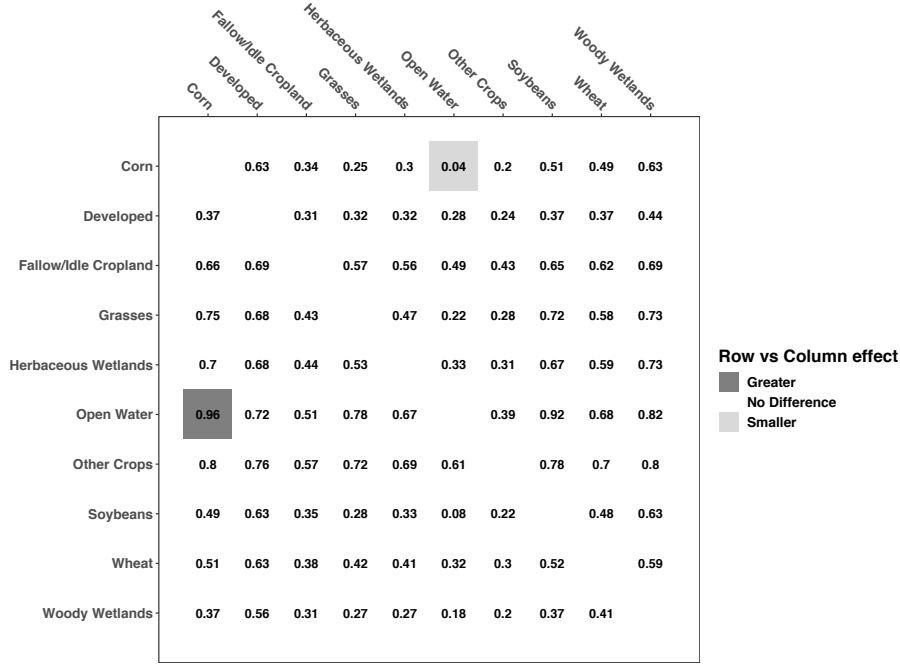

## S.6: Sensitivity of Estimates

We assessed the sensitivity of estimates to the choice of number of imputation data sets. The following figures show the 95% credible intervals from the posterior distribution based on 200 imputation data sets, 100 imputation data sets, and 1 data set corresponding to the most likely behavior classification (i.e., category with largest predicted probability based on the random forest). We see general consensus among the three scenarios with a general shrinking of estimates to zero with multiple imputation compared to most likely behaviors (Figure 3 and Figure 4). Credible interval widths for the imputation scenarios appear to be consistent or larger than the widths for coefficients estimated from the most likely behaviors.

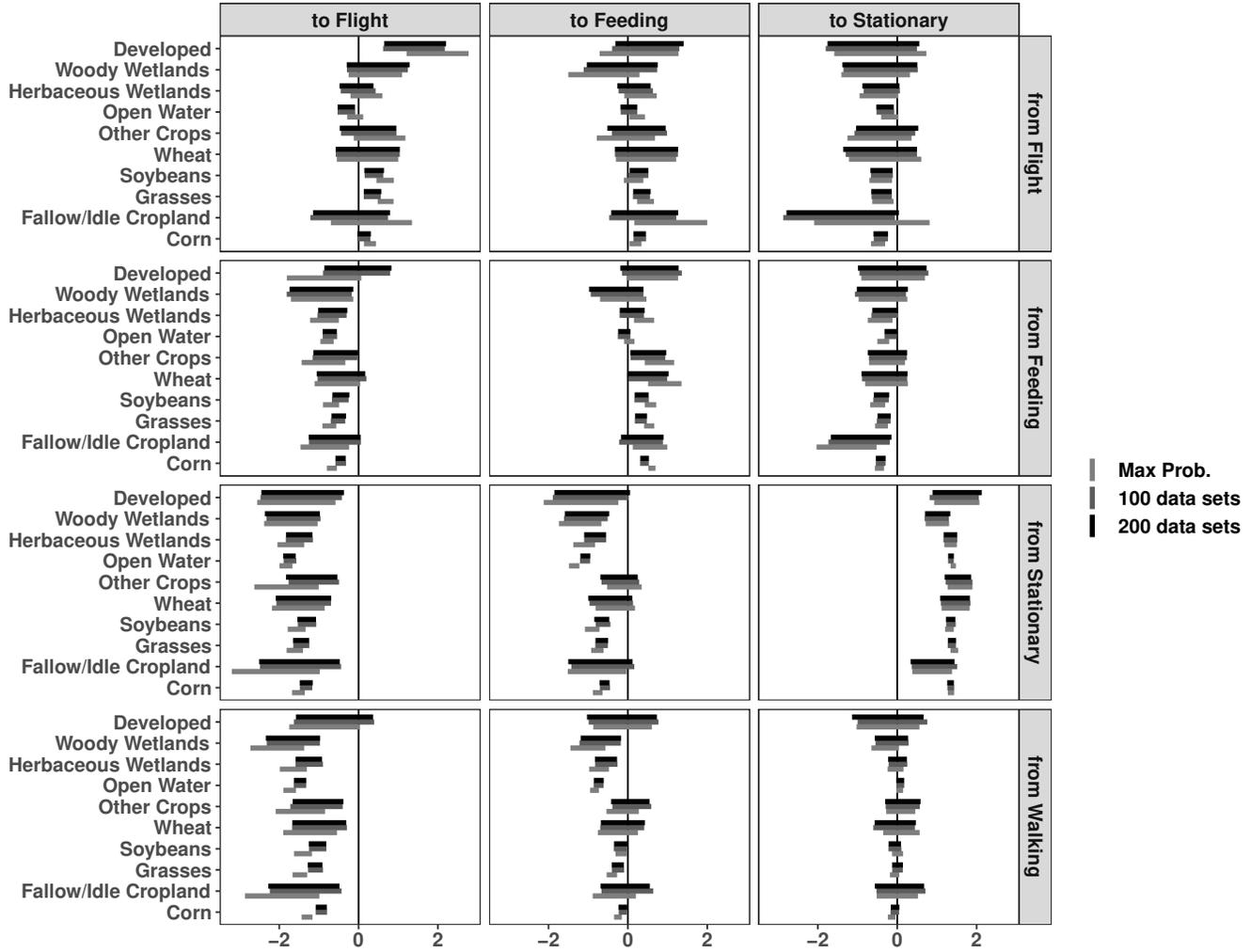

Figure 3: 95% credible intervals for habitat effects on the log-odds estimated in a Bayesian framework with either the most likely behavior class (Max Prob.), 100 imputation data sets, or 200 imputation data sets by behavior transition for six greater white-fronted geese in March 2018.

15

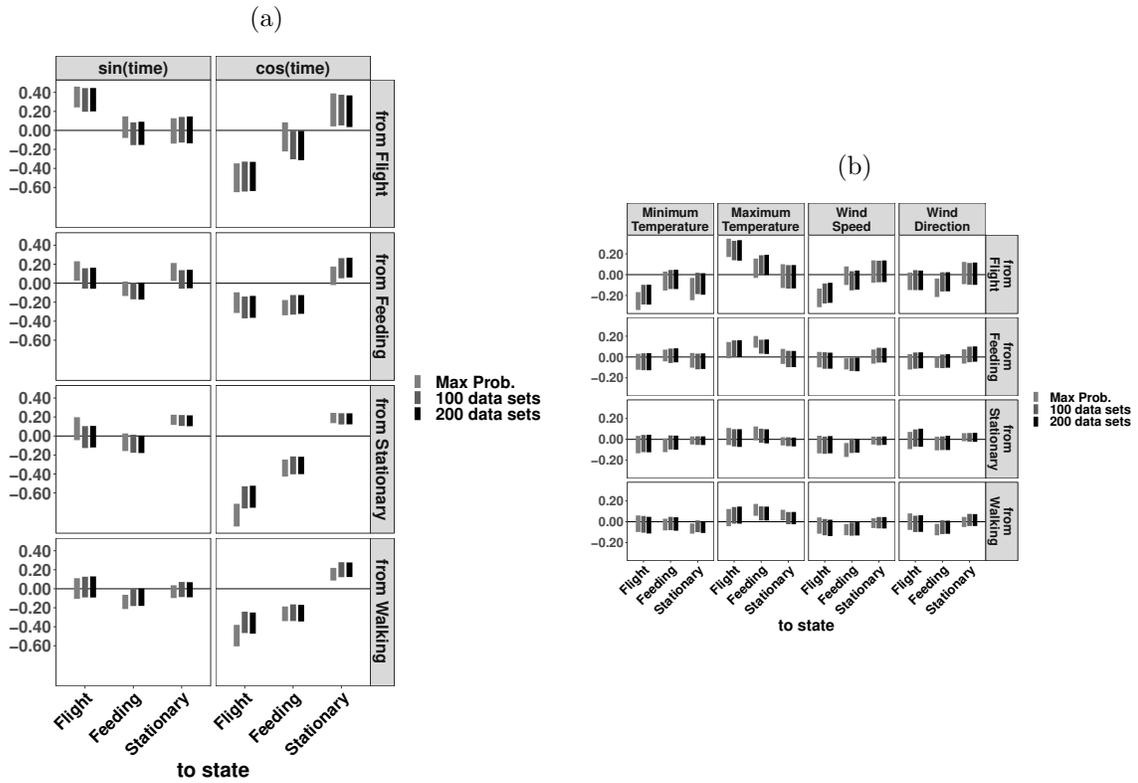

Figure 4: 95% credible intervals for diurnal (a) and weather (b) covariate effects on the log-odds estimated in a Bayesian framework with either the most likely behavior class (Max Prob.), 100 imputation data sets, or 200 imputation data sets by behavior transition for six greater white-fronted geese in March 2018.



\end{document}